\documentclass{cupconf}

  \checkfont{eurm10}
  \iffontfound
    \IfFileExists{upmath.sty}
      {\typeout{^^JFound AMS Euler Roman fonts on the system,
                   using the 'upmath' package.^^J}%
       \usepackage{upmath}}
      {\typeout{^^JFound AMS Euler Roman fonts on the system, but you
                   dont seem to have the}%
       \typeout{'upmath' package installed. cupconf.cls can take advantage
                 of these fonts,^^Jif you use 'upmath' package.^^J}%
      }
  \else
  \fi

  \checkfont{msam10}
  \iffontfound
    \IfFileExists{amssymb.sty}
      {\typeout{^^JFound AMS Symbol fonts on the system, using the
                'amssymb' package.^^J}%
       \usepackage{amssymb}%
         \let\leq=\leqslant
         
      }{}
  \fi

  \IfFileExists{amsbsy.sty}
    {\typeout{^^JFound the 'amsbsy' package on the system, using it.^^J}%
     \usepackage{amsbsy}}
    {}

\newsavebox{\astrutbox}
\sbox{\astrutbox}{\rule[-5pt]{0pt}{20pt}}

\newcommand\eg{e.g.,\ }
\newcommand\ie{i.e.,\ }
\newcommand\et{et~al.\ }
\newcommand\kms{km~s$^{-1}$}
\newcommand\lya{Ly\,$\alpha$}
\newcommand\civ{C\,{\sc iv}}
\newcommand\ciii{C\,{\sc iii]}}
\newcommand\mgii{Mg\,{\sc ii}}
\newcommand\feii{Fe\,{\sc ii}}
\newcommand\feiii{Fe\,{\sc iii}}
\newcommand\hb{H$\beta$}
\newcommand\oiii{[O\,{\sc iii]}}
\newcommand\mbh{$M_{\rm BH}$}
\newcommand{\lbol}{$L_{\rm Bol}$}
\newcommand{\lol}{$L_{\rm Bol}/L_{\rm Edd}$}
\newcommand{\Msol}{\mbox{$M_{\odot}$}}
\newcommand{\Msun}{\mbox{$M_{\odot}$}}
\newcommand{\lsim}{\stackrel{\scriptscriptstyle <}{\scriptstyle {}_\sim}\ }
\newcommand{\gsim}{\stackrel{\scriptscriptstyle >}{\scriptstyle {}_\sim}\ }

\usepackage{graphicx}

\title[Black-hole masses of distant quasars]{Black-hole masses of distant quasars}

\author[M.\ Vestergaard]
{M.\ns V\ls E\ls S\ls T\ls E\ls R\ls G\ls A\ls A\ls R\ls D$^{1,2}$}

\affiliation{$^1$
Steward Observatory, Department of Astronomy,
University of Arizona, 933 North Cherry Avenue, Tucson, AZ, 85718, USA\\[\affilskip]
$^2$ Current address:
Department of Physics and Astronomy, Tufts University,
Robinson Hall, Medford, MA 02155\\[\affilskip]}

\begin{document}

\maketitle

\begin{abstract}
A brief overview of the methods commonly used to determine or estimate the black hole mass in quiescent or active galaxies is presented and it is argued that the use of mass-scaling relations is both a reliable and the preferred method to apply to large samples of distant quasars.  The method uses spectroscopic measurements of a broad emission-line width and continuum luminosity and
currently has a statistical 1$\sigma$ uncertainty in the absolute mass values of about a factor of 4. Potentially, this accuracy can be improved in the future. When applied to large samples of distant quasars it is evident that the black hole masses are very large, of order 1 to 10 billion~\Msol, even at the highest redshifts of~4 to~6. The black holes must build up their mass very fast in the early universe. Yet they do not grow much larger than that: a maximum mass of $\sim$10$^{10}$~\Msol{}  is also observed. Preliminary mass functions of active black holes are presented for several quasar samples, including the Sloan Digital 
Sky Survey. Finally, common concerns related to the application of the mass scaling relations, \mbox{especially for high redshift quasars, are briefly discussed.}
\end{abstract}

\firstsection % if your document starts with a section,
\section{Introduction: Mass-estimation methods for active galaxies and quasars \label{intro.sec}}

The \textit{Hubble Space Telescope} has played a key role in our ability to detect supermassive black holes in the centers of nearby galaxies and to determine their mass through its high angular resolution. This has started a new, exciting era in which we are now able to study these massive objects and how they affect their environment and its cosmic evolution, and thereby get a more complete picture of structure formation and evolution over the history of the universe.

There are different ways in which to determine or estimate the black-hole 
mass in active nuclei, and they all differ somewhat from the methods applied 
to quiescent galaxies. Therefore, to place these methods in perspective and
to explain why the method of ``mass-scaling relations'' is preferred for distant 
active nuclei, I will start by giving a brief overview of the methods used to 
determine or estimate black-hole masses in quiescent and active galaxies.

In galaxies containing a quiescent or dormant central black hole, its mass is measured by determining the velocity dispersion of stars or gas close enough to the black hole that their dynamics are dominated by its gravity (\ie within the black hole's ``radius of influence'') by means of the virial theorem. While this appears the most direct way to measure the black-hole mass, this method cannot be applied to active black holes due to the strong glare from the central source. The most robust mass-determination method for active nuclei depends on its source type. Active galaxies are classified as Type~1 and Type~2, depending on whether the broad emission-line region can be viewed by us (Type~1) or not (Type~2). Type~1 sources tend to be oriented more face-on, and Type~2 sources are typically more highly inclined, such that gas and dust in the
host galaxy or in the outskirts of the central engine block our direct view of the broad-line region. For Type~2 sources, water mega-masers emitted by the molecular gas disk in the galaxy (\eg Miyoshi \et 1995) is a very reliable and accurate way to determine the black hole mass. For less-inclined Type~1 sources, the mega-maser method cannot be used. The most robust way to 
determine their black-hole masses is through the emission-line reverberation 
mapping method that relies entirely on the variability nature of the central  engine (\eg Peterson 1993; see also Peterson, this volume), and thus not on high angular-resolution observations as are needed for quiescent galaxies. At the present time, the monitoring data that has been collected allow the so-called one-dimensional reverberation mapping in which the broad emission-line time delay relative to continuum variations can be determined.  This delay is the light travel time of photons from the central continuum region and is, hence, the distance of the broad line gas from the continuum region. The black-hole mass is determined from the virial theorem by also measuring the velocity dispersion of this variable gas. Ultimately, we would like to do a two-dimensional analysis which also yields a velocity-delay map. With the latter, we would be able to determine the black-hole mass zero-point independently of other methods. The one-dimensional method does not allow this, so at present we are limited
to estimating the absolute-mass zero-point by other means. This is explained
next.

For quiescent local galaxies, the masses of their black holes display a strong, 
well-established correlation with the velocity dispersion of stars in the galactic bulge, far beyond the gravitational reach of the black hole (\eg Ferrarese \& Merritt 2000; Gebhardt et~al. 2000a), the so-called $M_{\rm BH}$--$\sigma_{\ast}$ relationship. Active black holes also display such a relationship (Gebhardt \et 2000b; Ferrarese \et 2001; Onken \et 2004), but given the unknown geometry and structure of the broad emission-line region, it is not yet possible to independently determine the zero-point of the relationship. However, if one assumes that the active black holes in the local neighborhood have essentially built up their mass, we can assume that the two $M_{\rm BH}$--$\sigma_{\ast}$ relationships are, in fact, one and the same, and thereby empirically establish the absolute zero-point offset of the mass scale of active black holes (\eg Onken \et 2004). The assumption is fair, since most of the active nuclei in the reverberation-mapping sample have black holes accreting at low levels, between 0.1\% to a few percent of the Eddington rate (Peterson \et 2004). 

It is not always possible to apply the primary methods of reverberation mapping
or mega-masers to determine the black hole mass. This is where ``secondary or
tertiary methods'' become useful. They typically build on or approximate the
primary methods. Table~1 gives an overview of these methods, which I name
``Secondary'' for convenience. For example, BL~Lac-type objects are viewed at such small inclination angles that the continuum emission is very strongly Doppler amplified, and the source variability reflects details of the powerful radio jets we are looking right into, rather than that of the continuum and line-emitting regions. Also, BL~Lacs typically have very weak or no broad emission lines, most likely due to the strong continuum boosting. For these types of sources, researchers have utilized the fact that they are hosted by elliptical galaxies. The physical parameters of elliptical galaxies (surface brightness $\Sigma_e$, effective radius, $r_e$, and stellar velocity dispersion $\sigma_{\ast}$) correlate such that they form a ``plane,'' the Fundamental Plane (Djorgovski \& 
Davis 1987; Dressler \et 1987). To determine the central black-hole mass, one can measure $\Sigma_e$ and $r_e$ so to infer $\sigma_{\ast}$ from the Fundamental Plane, and next use the $M_\mathrm{BH}$--$\sigma_{\ast}$ to infer the black-hole mass (\eg Woo \& Urry 2002; Falomo \et 2003). Unfortunately, owing in part to the finite thickness of the Fundamental Plane, the inference from this plane relationship itself can add as much as a factor of 4 uncertainty (Woo \& Urry 2002) to the mass estimates. This method can also be applied when only (host galaxy) imaging data are available. Alternatively, one can determine the luminosity of the galaxy (bulge) and from the $M_\mathrm{BH}$--$L$(Bulge) relationship (\eg Magorrian \et 1998), a sister relationship to the $M_\mathrm{BH}$--$\sigma_{\ast}$ relation, infer the black-hole mass (McLure \& Dunlop 2002). The lowest uncertainties that have been reported are a factor 3 to~4.  Unfortunately, this method is prone to larger typical uncertainties when applied to active galaxies, owing to the difficulties in separating the luminosity of the galaxy bulge from the strong nuclear source (\eg Wandel 2002). With spectroscopic measurements of the galaxy bulge $\sigma_{\ast}$, the black-hole mass can be inferred from the $M_\mathrm{BH}$--$\sigma_{\ast}$ relation with a significantly smaller uncertainty (Table~1). However, as the 
Ca\,\textsc{ii} triplet lines $\lambda \lambda$~8498, 8542, 8662 shift into the water vapor lines for $z \gsim\!\!0.068$, it becomes increasingly difficult and the measurement uncertainties increase (\eg Ferrarese \et 2001; Onken \et 2004).
Some authors have then resorted to determining $\sigma_{\ast}$ from a
collection of stellar features, although weaker, at shorter wavelengths (\eg Barth \et 2005). 

\begin{table}
\begin{center}
\begin{tabular}{lcccccc}
&\multicolumn{2}{c}{Low-$z$} &~~& High-$z$\\
\cline{2-3} \cline{5-5}\\[-8pt]
&Low-$L$~&High-$L$ &&High-$L$ &&Best Accuracy\\[1pt]
&LINERs, Sy\,2s~ &QSOs,\,Sy\,1s, BL Lacs&&QSOs&&(Dex)\\[5pt]
Scaling Relations &$\surd$ &$\surd$ &&$\surd$ &&0.5--0.6\\ % note $a$ \\
Via \mbh--$\sigma_{\ast}$:\\[3pt]
\hspace{0.2cm}
 $-$ $\sigma_{\ast}$ &$\surd$ &$\surd$ &&$\div$ &&0.3\\[3pt] % note $b$ \\[1pt]
\hspace{0.2cm}
 $-$ \oiii{} FWHM &$\surd$ &$\surd$ &&$\surd$ &&0.7\\[3pt] % note $c$ \\[1pt]
\hspace{0.2cm}
 $-$ Fundamental\\[1pt]
\hspace{0.2cm}
  Plane: $\sum_e, r_e$ &$\surd$ &$\surd$ &&$\div$ &&\llap{$>$}0.7 \\[5pt] % note $d$ \\[3pt]
Via \mbh{}--$L_\mathrm{bulge}$\\[1pt]
\hspace{0.2cm}
  \& scaling rel.:\\
\hspace{0.2cm}
 $-$ $M_R$  &$\surd$ &$\surd$ &&$\div$ &&0.6--0.7\\[2pt] % note $e$ \\[2pt]
%\multicolumn{6}{l}{{\sl Notes} $-$ ($a$) Improve $R$\,$-$\,$L$ relation; understand outliers, ($b$) Extend to more}\\
%\multicolumn{6}{l}{ distant AGNs and luminous quasars, ($c$) Understand scatter and outliers,}\\
%\multicolumn{6}{l}{($d$) Quantify and improve accuracy, ($e$) Calibrate to Reverberation masses}\\
\end{tabular}
\end{center}
\caption{Secondary mass-estimation methods}
\end{table}

Because reverberation mapping requires a lot of telescope time and resources,
this method is difficult or impossible to apply to large samples of sources, especially for distant objects (\eg Vestergaard 2004a; Kaspi 2001). It is also an issue that more luminous sources, owing to their larger size, vary on longer time scales and with smaller amplitudes, making reverberation-mapping analysis challenging. Even for the nearby PG quasars of order 10~years of monitoring data are required to constrain the black-hole mass well enough (\eg Kaspi 2001). Instead, one method that has been applied to obtain mass estimates of large samples makes use of the fact that the width of the \oiii\ $\lambda$5007 line approximates the velocity dispersion of the stars in the galaxy (Nelson 2000; Nelson \& Whittle 1996). The idea is to use the FWHM(\oiii) as a substitute for $\sigma_{\ast}$ and then the $M_{\rm BH}$--$\sigma_{\ast}$ relationship to infer the central mass.  However, the 1$\sigma$ statistical uncertainty in this method is a factor of 5 (Boroson 2003).

One other secondary method seems to perform better overall: the method sometimes referred to as ``mass-scaling relations.'' The method uses measurements of line width and continuum luminosity from a single spectrum of an active nucleus to estimate the black-hole mass. It is an approximation to the reverberation-mapping method, and therefore is also based on the virial theorem.
The method relies strongly on another result from reverberation mapping, namely
the radius--luminosity relationship, based on the size of the line-emitting region (as measured for a particular emission line) which can be estimated from the 
continuum luminosity. The advantages of this method are that it can both be applied to nearby and distant Type~1 active nuclei and to \mbox{large samples with} relative ease. The methods relying on accurate measurements of the host galaxy properties are limited to relatively low redshifts below $z$ of about 0.5 to 1.0,  since more distant host galaxies are difficult to characterize (\eg Kukula \et 2001). Mass-scaling relations presently provide an accuracy of the absolute-mass values of a factor of about 4. This is a method for which we can potentially even further improve the mass estimates in the future.

In this contribution, I will focus on mass-scaling relations as my preferred method to obtain mass estimates of black holes in large samples of distant
quasars: (a)~because of their easy application to a larger redshift range than the other methods, without going into the infrared regime; and (b)~due to the 
lower uncertainties associated. Before I discuss this method, I will address the key component of this method---namely the broad-line region radius--luminosity relationship (Section~\ref{rl.sec}).

Section~\ref{scalingrel.sec} is dedicated to mass~scaling relationships and important related issues. The distribution of quasar black-hole masses over the history of the universe is described in Section~\ref{Mdistrib.sec}, while the first mass functions based on various large quasar samples are presented in Section~\ref{MF.sec}. Potential issues with mass-scaling relations as commonly highlighted in the literature are discussed in Section~\ref{issues.sec}, before the summary and conclusions (Section~\ref{summary.sec}).

A flat cosmology with $H_0 = 70$~km~s$^{-1}/$Mpc and $\Omega_{\lambda}=0.7$ is used throughout.

\section{The radius--luminosity relationship \label{rl.sec}}

Peterson (this volume) has already introduced the radius--luminosity relationship: the relation between the characteristic size of the gaseous region emitting a particular broad emission line and the ionizing continuum luminosity that excites this gas. Therefore I will not discuss these relationships and their history in detail, but only emphasize a few issues that are important for their applications to estimating the black-hole mass for distant active nuclei. 

\subsection{The optical R--L relationship}

Peterson discussed the efforts to establish the physically relevant relationship between the size of the \hb-emitting region and the ionizing optical continuum luminosity without the contamination from stellar light in the host galaxy of the active nucleus (Bentz \et 2006a). Keep in mind the reason that the first $R$--$L$ relations published had a steeper slope is mostly due to the contamination from the host galaxy. The degree of contamination, even from nearby luminous quasars thought to entirely dominate their host-galaxy light, is in fact larger (Bentz \et 2006a) than originally assumed (e.g., Kaspi \et 2005). Therefore, the only radius--luminosity relation for $R$(\hb) and $L_{\lambda}$(5100~\AA) that should be applied is that of Bentz \et (2006a), for which the slope is firmly established to be 1/2, because this reflects the intrinsic, physical relationship.

\subsection{The UV R--L relationships}

Kaspi \et (2005) established that a similar R--L relationship exists for continuum luminosities in the UV and x-ray regions using reverberation results and archival data. In particular, $R$(\hb) scales with the monochromatic continuum luminosity at 1450~\AA\ as $R$(\hb) $\propto L_{\lambda}$(1450~\AA)$^{0.53}$.
Restframe UV luminosities are most useful when studying distant quasars, since we can then use ground-based observations in the observed optical region. 
However, in this case it is not the $R$(\hb)--$L$(UV) relationship that is relevant, but rather that pertaining to the UV emission line (\mgii\ or \civ) with which it is to be used (Section~\ref{scalingrel.sec}).  Prior to 2007, no such relationship could be established for \civ, mostly due to the unavailability of a sufficient data base (see Peterson, this volume). The inherent assumption in the mass estimates based on \civ\ at the time was therefore that the $R$(\civ)--$L$(UV) relationship that had to exist, would be similar to that of the $R$(\hb)--$L$(UV) relation (\eg Vestergaard \& Peterson 2006). The recent successes of Peterson \et (2005) to measure a \civ\ time delay (or lag) in the low-luminosity Seyfert galaxy NGC\,4395 and of Kaspi \et (2007) to measure a \civ\ lag for a luminous quasar has finally extended the luminosity range (to over 7 orders of magnitude in luminosity) of previously measured (reliable) \civ\ lags, allowing us to confirm that the slope of the $R$(\civ)--$L$(UV) relation (of 0.53) is entirely consistent with that for \hb.  This confirms the previous assumptions made and shows that, as it happens, the recently updated mass-scaling relationships based on \civ\ (Vestergaard \& Peterson 2006), which assumed $R$(\civ) $\propto L$(UV)$^{0.53}$, still hold without any modifications needed. 

It is worth noting that an $R$--$L$ relationship has not been established for the \mgii\ emission line, since only one reliable determination of the size of the \mgii-emitting regions exists at present (Metzroth \et 2006).  \mgii\ is the one emission line that has typically not been targeted in monitoring campaigns. Its isolation in the spectrum from the UV high-ionization lines and the optical Balmer lines has significantly contributed to this fact. The only reliable \mgii\ lag is based on improved reverberation data of NGC\,4151, which show that \mgii\ is consistent with being emitted from a distance twice that of \civ\ for two epochs;  yet within the uncertainties, the two lines are also consistent with being emitted from similar distances (Metzroth \et 2006). Apart from measurement uncertainties, the limitation that no contemporaneous measurements of the \hb\ and \mgii\ lags currently exists. However, new \hb-monitoring data of NGC\,4151 do suggest that \mgii\ and \hb\ are emitted at similar distances from the continuum regions (Bentz \et 2006b).

\subsection{Do the R--L relations apply to distant luminous quasars? \label{rlhiz.sec}}
There are a few reasons why we can comfortably apply these $R$--$L$ relationships to quasars beyond our local neighborhood in which they are defined. 
First, quasar spectra look remarkably similar (to first order) at all redshifts. This has been well demonstrated by Dietrich et~al.\ (2002), who show in their Figures~3 and 5 the strong similarity of the quasar broad-line spectra across the history of the universe when binning in either luminosity or redshift. While this complicates a thorough understanding of the broad-line region, it is a help in this case, because it shows that for all quasars the broad-line regions are essentially the same---whether you look at a quasar in your backyard or one at the furtherest reaches of the universe (e.g.,  Barth \et 2003; Jiang \et 2007). If the broad-line regions are similar, they will also obey the same radius--luminosity relationships. 
Second, a key point about the Kaspi \et (2007) results is that the quasar for which the size of the \civ-emitting region could be determined is located at a redshift of 3! Hence, these results also indicate that high-redshift quasars do have similar broad-line regions that respond to luminosity changes as do nearby active nuclei.  
Third, when we apply the $R$--$L$ relations to more distant quasars, we are, contrary what is sometimes stated, not extrapolating beyond the luminosity range for which these relations are defined. The $R$--$L$ relation pertaining to the optical continuum luminosity $L_{\lambda}$(5100~\AA) and $R$(\hb) is defined over 5 orders of magnitude in luminosity: from $10^{41}$~erg~s$^{-1}$ to $10^{46}$~erg~s$^{-1}$ (Bentz et~al.\ 2006a; see Figure~3 in Peterson's contribution in this volume). The UV $R$--$L$ relation, relevant for $R$(\civ) and $L_{\lambda}$(1350~\AA) or $L_{\lambda}$(1450~\AA)---these luminosities can be used interchangeably (Vestergaard \& Peterson 2006)---is defined over 7~orders of magnitude in luminosity: from $10^{39.3}$~erg~s$^{-1}$ to $10^{47}$~erg~s$^{-1}$ (Kaspi \et 2007). Next, I will compare these characteristic luminosity ranges to the luminosities of known quasars of various surveys.

% FIGURE 1
\begin{figure}
\centering
\includegraphics[height=6.56cm,width=6.56cm,angle=0]{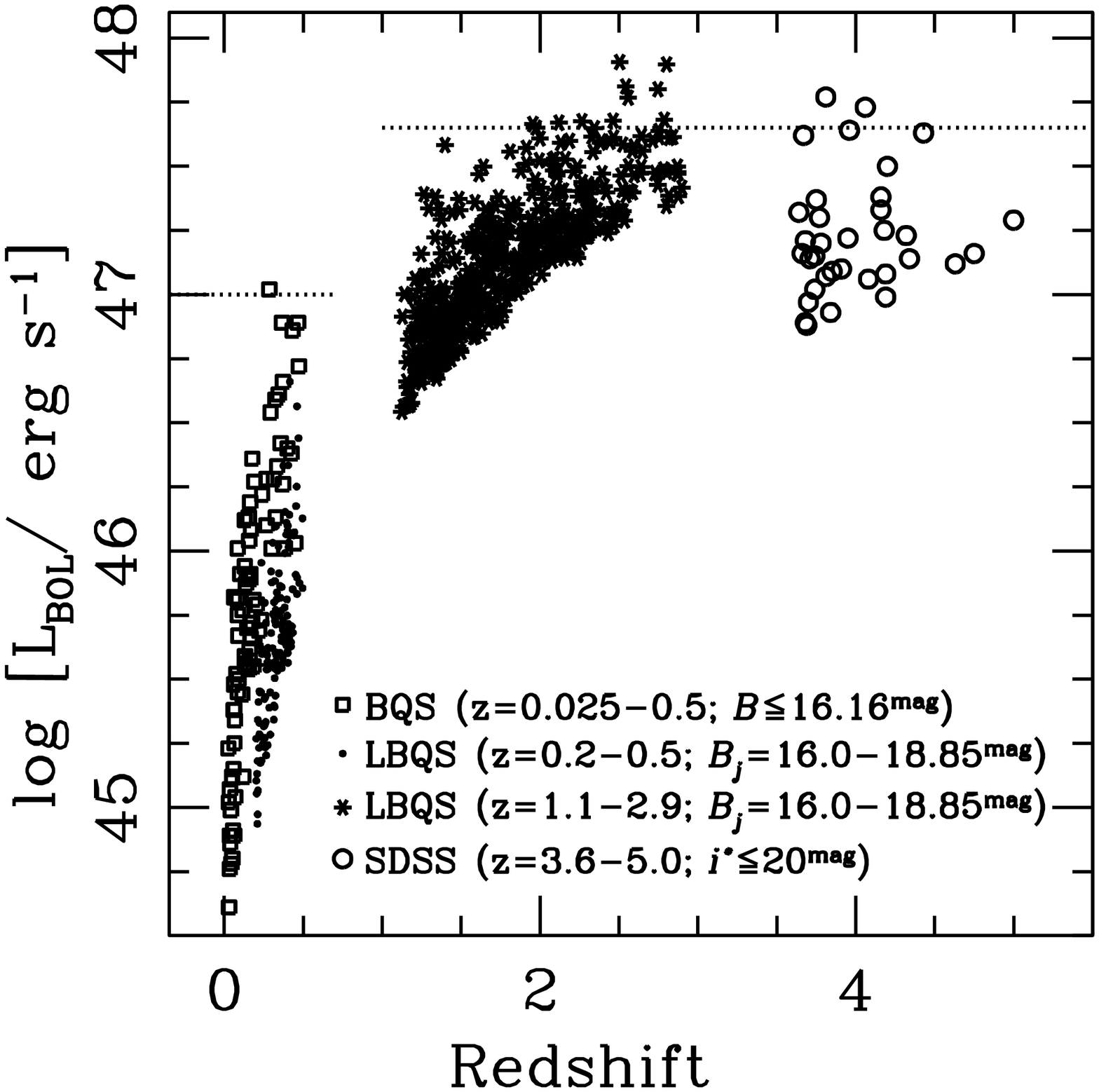}
\includegraphics[height=6.56cm,width=6.56cm,angle=0]{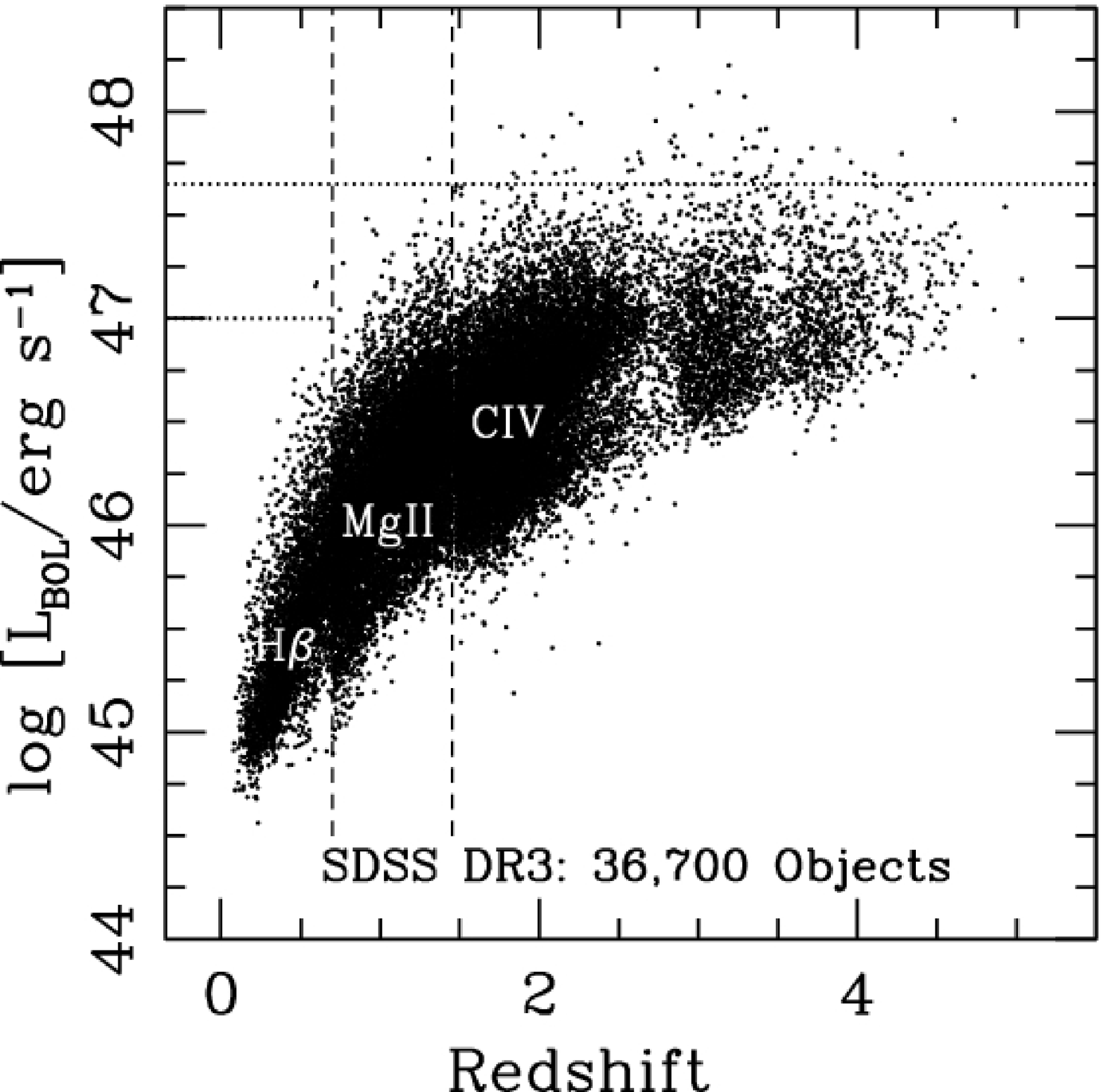}
\caption{Distribution of bolometric luminosities for various quasar samples (labeled). The dotted lines mark the maximum luminosity for which the optical (applied at $z \lsim\!\!0.9$) and UV $R$--$L$ relationships are defined. Very few quasars, even at high~$z$, have luminosities exceeding these, so the use of the $R$--$L$ relations is reasonable. Left: The Bright Quasar Survey (BQS), the Large Bright Quasar Survey (LBQS), and the color-selected sample presented by Fan et~al.\ (2001), as labeled. Right: The Sloan Digital Sky Survey Data Release~3 quasar catalog. Vertical dashed lines indicate the subsets where different
emission lines (labeled) are used for the mass estimates (left to right: \hb, \mgii, \civ).
 }\label{fig:Lz}
\end{figure}

Figure~\ref{fig:Lz} shows the luminosity distributions of a few quasar samples, including the Bright Quasar Survey (BQS) at $z \leq 0.5$, the Large Bright Quasar Survey (shown in two redshift ranges: $z \leq 0.5$ and $1.1 \leq z \leq 2.9$ that are discussed in later Sections), the color-selected sample from the Sloan Digital Sky Survey Fall Equatorial Stripe (Fan et~al.\ 2001), and the SDSS Data Release~3 quasar catalog (Schneider \et 2005). These samples are discussed further in Sections \ref{Mdistrib.sec} and \ref{MF.sec}. I converted the monochromatic power-law continuum luminosities to bolometric luminosities, \lbol, in order to show the luminosities of the individual samples in the same diagrams. The continuum luminosities are determined through a decomposition of the quasar spectra into a power-law continuum component, an iron-emission component using the best and most complete iron templates in the UV (Vestergaard \& Wilkes 2001) and the optical (Veron \et 2004), and a line-emission component. The \lbol\ values are estimated by a simple scaling factor applied to the monochromatic luminosities of 4.62, 4.65, 5.8, 9.74, and 10.5 
(Vestergaard 2004b, and based on Richards et al.\ 2006) for the $L_{\lambda}$(1350~\AA), $L_{\lambda}$(1450~\AA), $L_{\lambda}$(3000~\AA), $L_{\lambda}$(4400~\AA), and $L_{\lambda}$(5100~\AA) luminosities, respectively. 

The luminosity distributions in Figure~\ref{fig:Lz} show that, as expected, more distant quasars are also more luminous. The lower flux limits of the different surveys uncovering these quasar samples are evident by the sometimes sharp cutoff of the distribution toward lower luminosities. In each panel I have indicated the maximum luminosity, in \lbol\ units, for which the $R$--$L$  relationships are defined for the optical continuum luminosity, $L_{\lambda}$(5100~\AA), and for the UV continuum luminosities, $L_{\lambda}$(1350~\AA) and $L_{\lambda}$(1450~\AA). They are $10^{47}$ and $10^{47.65}$~erg~s$^{_1}$, respectively, given the bolometric scaling factors introduced above. It is clear from Figure~\ref{fig:Lz} that very few quasars, even the distant, most luminous ones, have luminosities beyond the luminosity ranges where the $R$--$L$ relations are defined. This holds both at low and high redshift. Statements that the mass-scaling relationships may not be valid for distant luminous quasars because of a need to extrapolate far beyond the validity of the $R$--$L$  relations for these sources are clearly not supported.

\subsection{The power of the R--L relationships}

The $R$--$L$ relationships have tremendous power---providing they are reliable, of course---as discussed here and by Peterson (this volume), we have good reasons to believe that they are. They allow us to determine black-hole mass estimates of thousands of quasars on a relatively short timescale from a single spectrum of each source. This is in sharp contrast to the large amount of time and
resources required to determine the mass using reverberation mapping. However,
$R$--$L$ relationships do not marginalize monitoring campaigns and the reverberation method. In fact, reverberation mapping is still very important,  because it provides the time lags and sizes of the various line-emitting regions and the more directly and more robustly determined masses to which the scaling relations are anchored.

\section{Single-epoch mass estimates based on scaling relations \label{scalingrel.sec}}

Scaling relations are approximations to the mass estimates based on the  reverberation mapping method, as discussed by Peterson in this proceedings. Thus, similar to reverberation-based masses, the single-epoch mass estimates rely on the virial theorem, where $M_\mathrm{BH} = V^2 R/G$, $G$ is the gravitational constant and $R$ is the distance from the black hole to the gas with 
velocity dispersion, $V$. It is thus assumed that the broad-line emitting gas that is used to probe the black-hole mass is gravitationally dominated by the black-hole potential.  It has by now been firmly established to hold for most, if not all, active galaxy broad-line emitting gas (Peterson, this volume; Peterson \& Wandel 1999, 2000; Onken \& Peterson 2002; Metzroth \et 2006). 

Specifically, these (mass-) scaling relations use the broad-line width as a measure of proxy for the velocity dispersion of the broad emission line gas.  The line width is parameterized either as the line dispersion $\sigma_\mathrm{line}$ (the second moment of the line profile) or the full-width-at-half-maximum, FWHM.  As discussed later and by Peterson \et (2004) and Collin \et (2006), the line dispersion is the preferred line-width measure for its robustness. The continuum luminosity is used to estimate the distance to the line-emitting gas via the important $R$--$L$ relationship introduced in Section~\ref{rl.sec}, since, clearly, no time delays can be determined without extensive multi-epoch observations. In summary, the mass-scaling relations take the form of either:
\[
M_\mathrm{BH} = \mathrm{C1}\times \mathrm{FWHM(line)}^2 \times L_{\lambda}^{\beta}~~,
\]
or 
\[
M_\mathrm{BH} = \mathrm{C2}\times \sigma_\mathrm{line}^2 \times L_{\lambda}^{\beta}~~,
\] 
where C1 and C2 are the mass zero-points for each of the equations and the index $\beta$ is essentially 0.5 for continuum luminosities (Section~\ref{rl.sec}). The specifics of the most recent updates to these relationships for \civ\ and \hb\ are presented by Vestergaard \& Peterson (2006), to which the interested reader is referred. 

The possibility of using measurements from a single observation of an active nucleus to estimate the mass of its central black hole was first introduced by Wandel, Peterson, \& Malkan (1999) in the optical wavelength region using the \hb\ emission line and the ionizing luminosity estimated from photoionization theory. Vestergaard (2002) expanded this into the UV-wavelength region using
the \civ\ emission line by calibrating single-epoch measurements of FWHM(\civ) and $L$(1350~\AA) to black hole masses obtained with reverberation mapping data. This was followed by work using FWHM(\civ) and $L$(1450~\AA; Warner \et 2003) and FWHM(\mgii) and $L$(3000~\AA; McLure \& Jarvis 2002). In the latter study, \mgii\ is assumed to be emitted from the same distance as \hb,
such that the \mgii\ FWHM is substituting the \hb\ line width.  With the
availability of UV-scaling relations, it became more easily possible to estimate black-hole masses of distant quasars residing at the highest observable redshifts, and for large samples thereof. 

\subsection{An empirical determination of the f-factor \label{f.sec}}

In 2004, the reverberation-mapping database underwent a major revision 
(Peterson et~al.\ 2004), which resulted in improved determinations of black-hole mass and the radius of the variable \hb-emitting gas for the 36 active nuclei monitored at that time. As summarized in that paper (and Peterson, this volume), an $f$-factor enters the virial mass estimates owing to the unknown geometry and structure of the broad emission-line region. The appropriate value of $f$ is largely unknown, but is expected to be of order unity. However, Onken et~al.\ (2004) empirically determined the average value of the $f$-factor to be 5.5 for the reverberation sample of active nuclei. For those sources with reliable (and completely independent) determinations of both black-hole mass and host-galaxy bulge stellar-velocity dispersion, Onken et~al.\ made the assumption that intrinsically the $M$--$\sigma$ relationships for active and quiescent galaxies are the same. The distribution of black-hole masses and stellar-velocity dispersion for the active galaxies show the same slope, but an offset in the zero-point, suggesting that the assumed $f$-factor at the time was not representative. Prior to 2004, we had to make assumptions about the value of $f$ so to get an estimate of the black-hole mass.  The most simple kinematical structure, an isothermal gas distribution, with equal gas velocities in all directions, was often adopted, yielding an $f$ value of 0.75. It is, in fact, a very important point that the first values of $f$ adopted were entirely \textit{arbitrary}, but were accepted due to the lack of a more qualified guess at this time. The work by Onken et~al.\ (2004) has made all other previously assumed values of $f$ obsolete. It is worth emphasizing, since this is often misunderstood, that this 
empirical determination of $f=5.5$ is valid for random inclinations of the broad emission-line region with respect to our line of sight. The average value of $f$ was determined such that some sources will have the mass either overestimated or underestimated, but on average for the reverberation sample, the mass zero-point is consistent with the $M$--$\sigma$ relationship for quiescent galaxies. I will discuss in Section~\ref{inclination.sec} that the effects of source inclination can be minimized simply by using the line dispersion to measure the line width, or by making an $f$-factor correction to the mass estimates based on the FWHM. 

It is important to note that if one seeks to determine the black-hole mass of a \emph{single} object based on a single-epoch spectrum---as opposed to the masses for a sample of objects for which the mass-scaling relationships 
discussed in Section~\ref{newMrel.sec} can be used---then one should use $f=3.85$ along with the line dispersion to determine the mass (Collin \et 2006). Since a single-epoch spectrum resembles the mean spectrum much more than the RMS spectrum, and given that the mean and rms profiles are often different, the appropriate $f$-factor is also different. In other words, to recover the reverberation black-hole mass of one of the reverberation-mapped active galaxies using the line dispersion measured from a single spectrum, $f=3.85$ is needed. But, if only the FWHM of the line is available, one should use Eq.~7 of Collin \et (2006) to determine the appropriate $f$-factor, which depends on the FWHM value.

\subsection{Improved mass-scaling relations for H$\beta$ and C\,\textsc{iv} \label{newMrel.sec}}

Since the Peterson et~al.\ (2004) and Onken et al.\ (2004) papers significantly revised the reverberation data on which the Vestergaard (2002) scaling relations are anchored, a revision of the latter was necessary.  Using a larger database than the 2002 calibrations, Vestergaard \& Peterson (2006) present updated and improved scaling relationships for both the \civ\ and \hb\ emission lines, calibrated
to the revised reverberation-mapping masses.  Equations using either the FWHM or line dispersion are presented for \civ. Note that since the scaling relationships are calibrated directly to the reverberation masses---deemed the most accurate mass determinations of the objects in the reverberation-mapping sample---the $f$-factor is absorbed in the zero-points of the mass estimates. The Vestergaard \& Peterson study shows that, although the scatter of the single-epoch mass
estimates relative to the reverberation masses has decreased somewhat owing to the improved data base used, there is still some scatter remaining, amounting to a 1$\sigma$ statistical uncertainty in the absolute values of the mass estimates of a factor 3.5 to 5, depending on the emission-line and line-width measure. This includes the estimated uncertainty in the absolute zero-point of the reverberation masses of a factor of 2.9 or less, as estimated by Onken \et (2004). This means that there is a 68\% chance that the mass estimate of a single object is accurate to within a factor 3.5--5, but there is a 95\% chance it is accurate to within a factor of about 6--7. 

The remaining scatter is possibly due to several factors. For one, the intrinsic scatter in the $R$--$L$ relationship dominates the uncertainties in the mass estimates. That scatter is likely related to the detailed geometric and kinematical structure of the broad-line region that we are unable to account for. In addition, active nuclei do vary in luminosity. The $R$--$L$ and virial relationships indicate that as the nuclear luminosity changes, so will the size of the line-emitting region, and its location will dictate the gas velocity for a given black-hole mass. As a result, the $R$ and $V$ values should slide along the virial relationship, 
$R \propto V^{-1/2}$, as the luminosity changes. This has been clearly demonstrated for the Seyfert~1 galaxy, NGC\,5548 (Peterson \& Wandel 1999, 2000) and for several broad emission lines. And the mass estimates should always be consistent. However, deviations from a tight relationship may be expected in a couple of situations. For example, if the broad-line region has not yet adjusted to the luminosity change such that the actual $R$ and the measured $V$ may not accurately reflect the measured $L$. With BLR sizes of hundreds of days for luminous quasars (\eg Kaspi \et 2000, 2007) this can perhaps never be avoided. Another possibility for deviation is when the density distribution of the gas in the broad-line region is not smooth or ``typical'' (\ie
based on the $R$--$L$ relation, which yields an ensemble average). The different emission lines could be emitted from regions lying closer or farther from one another than what is ``typical.'' According to the Locally Optimized Cloud model of the broad-line region (Baldwin \et 1995), the predominant part of the flux in a particular emission line is emitted from the part of the broad-line gas that is most efficient, at a given time, of emitting this line.  The different emission lines will thus be emitted from different subregions in the broad-line region determined by the detailed physical conditions, such as the gas density and the incident photon flux. And since the broad~line region is a dynamic place, the locations of most efficient \civ-line emission relative to that of \hb\ may change in time and as the changing luminosity allows us to ``probe'' different part of the broad-line region. In other words, there may always be some scatter appearing in the $R$--$L$ relationships, and thus always some intrinsic uncertainties in the mass estimates that will limit how well we can establish the black-hole mass, especially for individual objects. However, the hope is that as we learn more about the broad-line region, we are able to make appropriate adjustments to these relationships, so the accuracy of the black hole mass estimates and of distant quasars will improve.

One additional factor that can affect mass estimates is the relative source inclination of the broad-line region to our line of sight. I discuss this separately in Section~\ref{inclination.sec}.

Unfortunately, the Vestergaard \& Peterson calibrations were published before Collin et~al.\ (2006) demonstrated that the \hb\ FWHM line widths may, in fact, \mbox{provide a biased} estimate of the broad-line gas velocity dispersion, depending on the inclination of the active nucleus with respect to our line of sight. So, while the Vestergaard \& Peterson scaling relations are anchored in the reverberation masses for which inclination effects are reduced, the relations based on FWHM do not directly take into account the possible inclination effects that can affect the measured FWHM values.  This issue, for both \hb- and \civ-scaling relations, will be addressed in a forthcoming paper.

\subsection{Mass-scaling relations for Mg\,\textsc{ii}}

The existing relations involving the \mgii\ emission line presented by McLure \& Jarvis (2002) can no longer be used because: (a)~these mass estimates turn out to be inconsistent with those based on \hb\ or \civ\ by up to a factor of 5 (Dietrich \& Hamann 2004), and (b)~the relationship has not been updated to the new reverberation-mass scale. I have instead established new relationships using a high signal-to-noise subset of SDSS quasars for which \mgii\ and one other broad emission line (\hb\ or \civ) is also present. These new relations make no prior assumptions of where \mgii\ is emitted. The relationships and the details of this analysis will be presented in a forthcoming paper (Vestergaard \et 2007, in preparation).

\subsection{Relationship involving the \hb\ luminosity}

For sources where the luminosity at 5100~\AA\ is not representative of the nuclear luminosity due to contamination from the host-galaxy light or the high-energy tail of the powerful radio emission in radio-loud sources that cannot easily be corrected, the \hb-line luminosity can be used to approximate the size of the \hb-emitting region (\eg Wu et~al.\ 2004).  This is valid because the \hb-line flux is directly proportional to the ionizing luminosity, which is the determining factor for the size of the broad-line region. Notably, since the \civ-emission line is partially collisionally excited, this assumption is not valid for \civ. 
This is also emphasized by the Baldwin Effect (an inverse correlation between the \civ\ equivalent width and the continuum luminosity): the continuum and \civ-line luminosities do not scale with one another. However, there is also no real need for an alternate luminosity measure---since neither the host-galaxy light, the high-energy tail of the radio emission, or even the UV iron emission between \lya\ and \civ\ make a significant contribution to the 1350~\AA\ or 1450~\AA\ continuum luminosity (\eg Elvis \et 1994). A recalibration of the relationship involving the \hb-line width and \hb-line luminosity is presented by Vestergaard \& Peterson (2006).

This relationship is not applied to the quasar samples discussed in this contribution.

\section{Black-hole masses in distant quasars \label{Mdistrib.sec}}

\subsection{Quasar samples \label{Qsamples.sec}}

In this Section I discuss the results of applying the mass-scaling relations to large samples of quasars: namely, the Bright Quasar Survey (BQS; Schmidt \&
Green 1983), the Large Bright Quasar Survey (LBQS; Hewett \et 1995), the Sloan Digital Sky Survey (SDSS) Data Release~3 quasar catalog (Schneider \et 2005), the color-selected sample of Fan \et (2001) based on the SDSS Fall Equatorial Stripe, the $z \approx 2$ sample of radio-loud and radio-quiet quasars
studied by Vestergaard \et (2000) and Vestergaard (2000, 2003, 2004b), and
several $z \approx 4$ samples from the literature. These samples and the data, with exception of the LBQS and the SDSS DR3 Quasar samples, are discussed in detail by Vestergaard (2004b). Details of the analysis of the LBQS and SDSS DR3 samples will be presented elsewhere (Vestergaard et~al.\ 2007, in preparation).  However, it is appropriate to point out that the spectral measurements are made after modeling and subtraction of the \feii\ and \feiii\ emission and, in the case of the DR3 sources below $z = 0.5$, the stellar light from the host galaxy; the
LBQS quasars were selected to be strictly point sources, which limits the contribution from the host galaxy. The \mgii\ and \hb\ lines, in particular, are embedded in broad bands of iron emission, which can be very strong in some sources. If the iron emission is not (well) fitted and subtracted, it will bias the line-width measurements. In particular, for \mgii\ one will systematically obtain smaller line widths because half of the broad \mgii-emission-line flux is blended with the iron emission (Vestergaard \& Wilkes 2001). To minimize measurement errors due to noise, bad pixels, and narrow absorption, the line-width measurements were made on model fits to the broad emission line that are representative of the broad profiles.

The black-hole mass estimates are based on the updated calibrations of Vestergaard \& Peterson (2006) for \hb\ and \civ; this work also lists the most current mass determinations and estimates of the BQS sample. For \mgii, the new relations discussed in Section~\ref{scalingrel.sec} were applied.
I am using the scaling relations based on the FWHM line widths, because most of the data are not of sufficient quality to allow the line dispersion to be significantly measured.  Since the line dispersion is an integral measure, it is less affected by (residual) narrow-line emission, but is much more affected by flux in the extreme wings of the line profiles, as well as the placement of the underlying
continuum level.  For these data that are not of very high signal-to-noise, contrary to reverberation-mapping data, the use of the FWHM allows us to probe a much larger subset of the quasar samples. As discussed in Section~\ref{inclination.sec}, corrections can be applied to the mass estimates based on FWHM measurements to reduce potential effects of source inclination. While this is the best approach, it has not been done in the current work (for the sake of consistency), because a correction only exists for the \hb\ line and not yet for the \civ\ and \mgii\ lines.

% FIGURE 2
\begin{figure}
\vbox{
\hbox{
\includegraphics[height=6.5cm,width=6.5cm,angle=0]{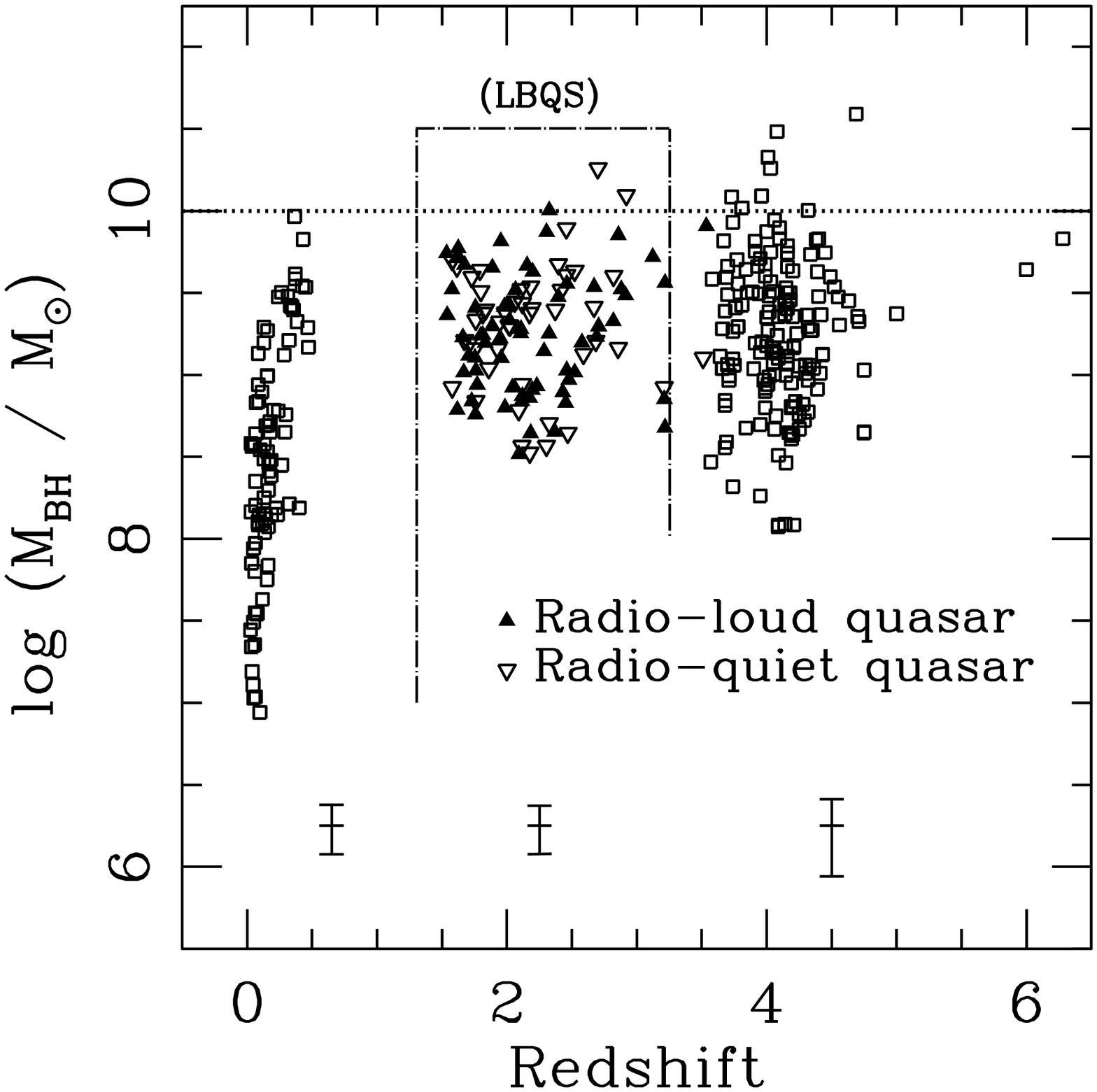}~
\includegraphics[height=6.5cm,width=6.5cm,angle=0]{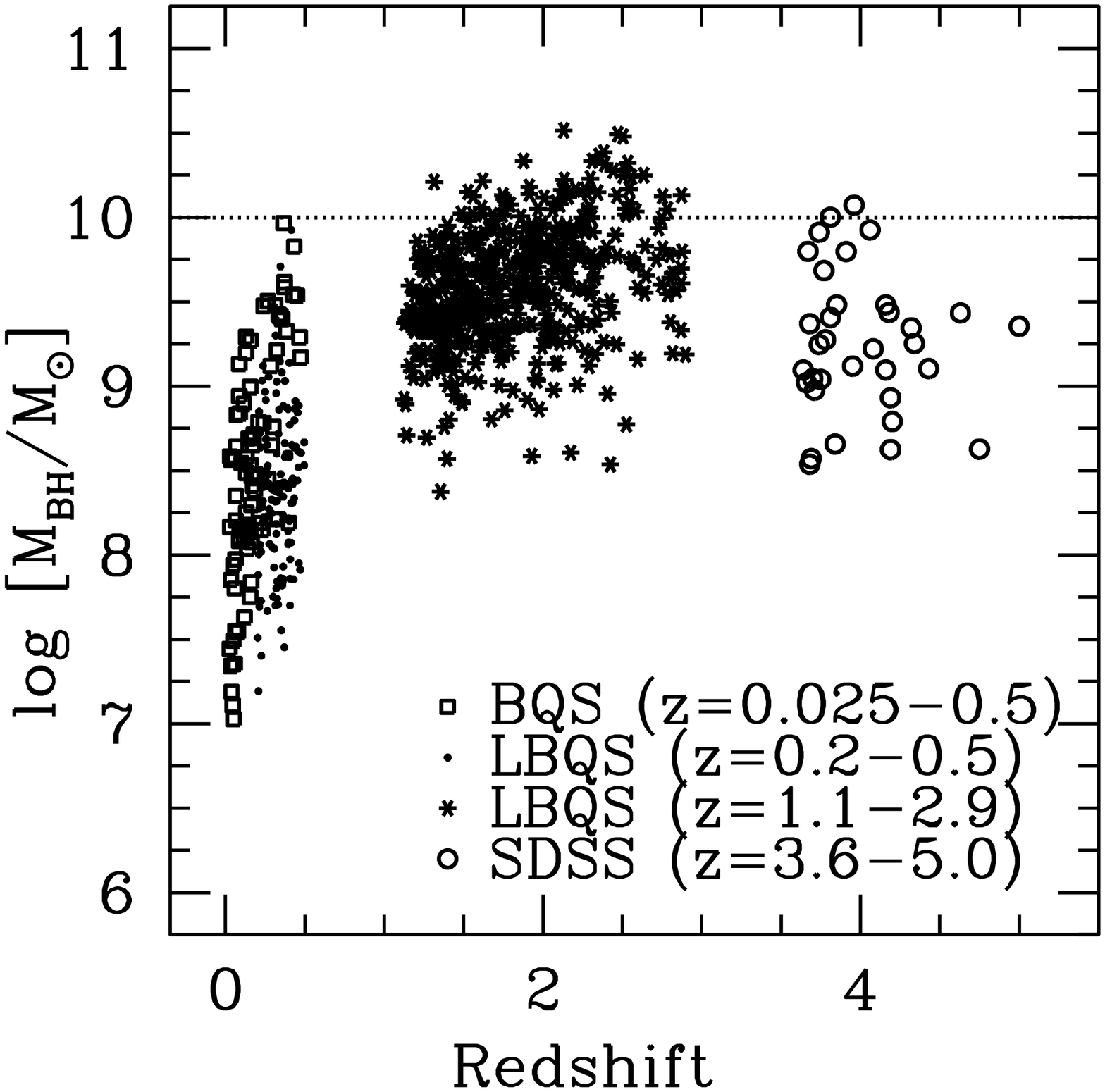}
}
\centering
\includegraphics[height=6.5cm,width=6.5cm,angle=0]{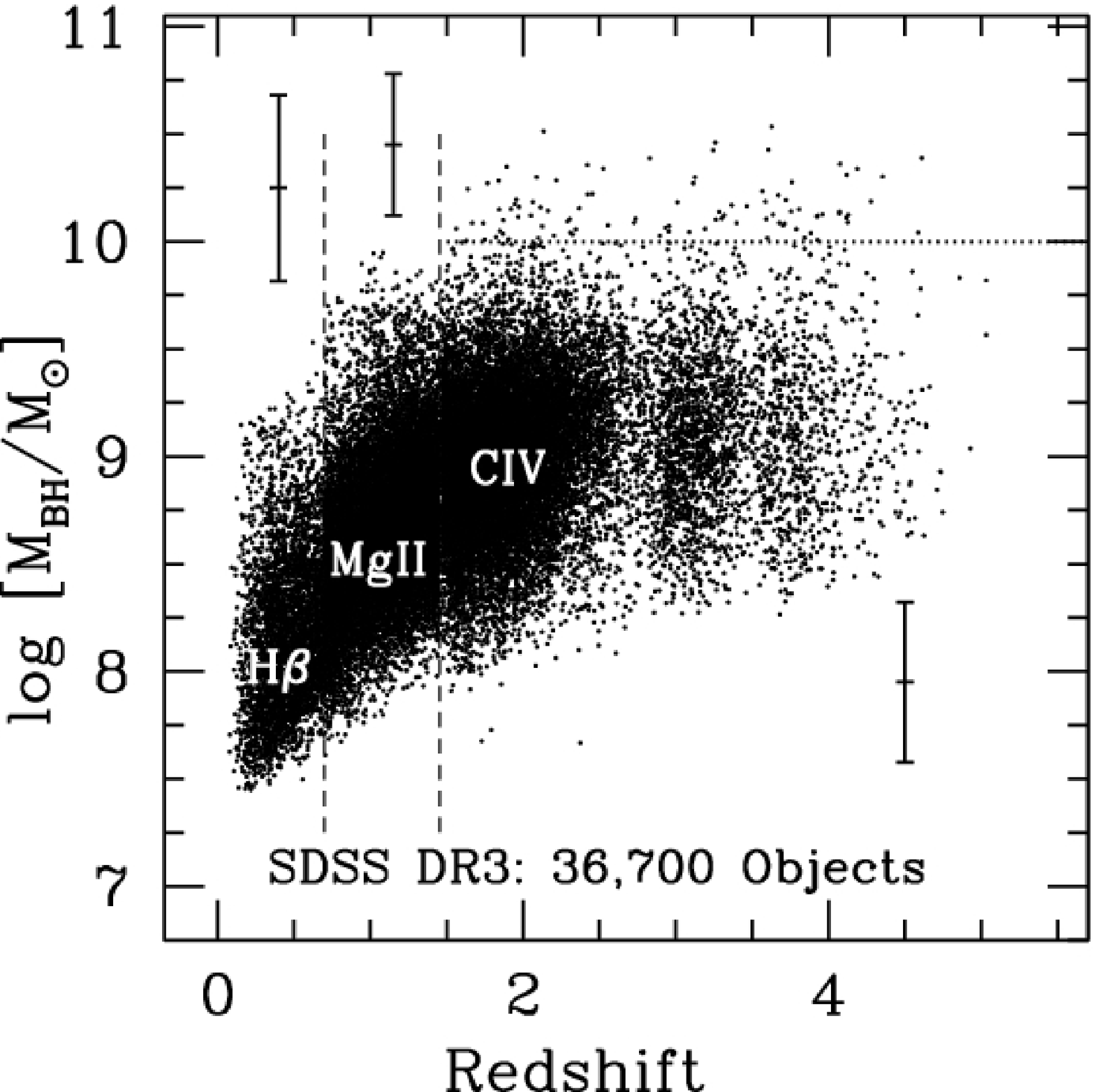}
}
\caption{Distribution of black-hole mass estimates for different quasar samples. These mass estimates are based on the Vestergaard \& Peterson (2006) single-epoch scaling relations and are computed in a flat cosmology with $H_0 = 70$~km~s$^{-1}$/Mpc and $\Omega_{\lambda}=0.7$. A mass of $10^{10}$\Msun\ is indicated by the dotted line. Upper left panel: The Bright Quasar Survey (BQS), the intermediate redshift sample studied by Vestergaard et~al.\ (2000), Vestergaard (2003, 2004b), and various samples of $z \gsim 4$ samples from the literature studied by Vestergaard (2004b).  The measurement uncertainty (not including the statistical accuracy of the mass estimates) is indicated at the bottom of the panel. Upper right panel: The BQS, The Large Bright Quasar Survey, and the Color-Selected Sample in the Fall Equatorial Stripe (Fan et~al.\ 2001). Lower panel: The Sloan Digital Sky Survey Data Release~3 Quasar Catalog. Vertical dashed lines indicate the subsets where different emission lines (labeled) are used for the mass estimates (left to right: \hb, \mgii, \civ). The error bars in each region indicate the typical uncertainty in the mass estimates and includes the statistical uncertainty in the mass zeropoint of about a factor of 4. The decreased densities of objects at redshifts of about 2.8 and 3.5 is due to the increased difficulty in selecting quasars at these redshifts, as in color space the quasar tracks overlap with the stellar locus. 
 }\label{fig:Mz}
\end{figure}

\subsection{Does radio-loudness depend on black-hole mass?}
The sample selection of the $z \approx2$ quasar sample makes it an ideal
dataset for studying the properties of quasars with respect to radio-loudness,  
since each individual radio-loud quasar is matched in redshift and luminosity 
to a radio-quiet quasar in the sample.  The distribution of masses in 
Figure~\ref{fig:Mz} shows that when luminosity and redshift differences are
eliminated, there is no difference in black-hole mass for quasars of different
radio-loudness, as the BQS otherwise suggests (Laor 2000); that result
is probably due to selection effects.

\subsection{Distributions with redshift \label{zdistrib.sec}}

The distributions of the black-hole mass estimates for the different quasar samples discussed in Section~\ref{Qsamples.sec} are shown in Figure~\ref{fig:Mz}. The distributions at the low-mass end are strongly affected by the flux
limits of the surveys by which the quasars were found. At low redshift, say $z<0.5$, we can probe less luminous sources than at higher redshifts. Since the BQS is a ``bright'' quasar sample in the nearby universe, it has black-hole masses spanning 3 orders of magnitudes. It contains some quite luminous sources like 3C\,273, which has a black-hole mass of a few billion solar masses. 
At redshifts higher than 1, we can typically only probe down to a few times
$10^8$~\Msol. The lack of data points below this mass is therefore an artifact of our sample selection. We do expect the quasar population to extend much below the current flux limit cutoff. 

It is characteristic that for the quasars we do detect at high redshift, say at $z \gsim\!\!1$, their black-hole masses are typically very large---of order a billion solar masses or more. This holds even beyond a redshift of 3 to 3.5, at which the quasar space density is declining, and for the most distant quasars known at $z \gsim\!\!6$. This shows that black holes must build up their mass very quickly, since, for example, at a redshift of 6.3 the universe was less than 900 Myrs old. The first star formation is thought to occur at redshifts between 6 and 9, based on the presence of heavy elements in the spectra of quasars at $z \approx 5$ (Dietrich \et 2003a,b) and $z \approx 6$ (\eg Barth \et 2003; Maiolino \et 2005; Jiang \et 2007; Dwek \et 2007) and chemical evolution models (\eg Fria\c{c}a \& Terlevich 1998; Matteucci \& Recchi 2001). While the black holes may form earlier, this does show that given the short time available, supermassive black holes must evolve rapidly. 

It is noteworthy that there appears to be a maximum black-hole mass for quasars at 10~billion solar masses. This mass value is marked in the diagrams of Figure~\ref{fig:Mz} for reference. There are very few black holes with masses above $10^{10}$~\Msol, even for very large samples like the SDSS. While SDSS does have an upper flux limit to the survey, these limits do not affect the quasar selection at high redshift. The number of objects with mass estimates above $10^{10}$~\Msol{} are statistically consistent with the uncertainties in the black-hole mass estimates. Therefore it is not unreasonable to conclude that there appears to be a limit to how massive actively accreting black holes can become, and that this limit lies at or near $10^{10}$~\Msol. At this point, it is unclear whether this limit is due to the dark matter halo in which the black hole resides, whether it has to do with details of the accretion physics, or whether the black hole is simply exhausting its accessible fuel supply. Of course, it could be due to a combination of all three possibilities. 

The distributions indicate that there is a decline in black-hole mass from a redshift of about 1.5 to the present. The SDSS survey does have a bright flux limit that deselects the brightest (and hence the most luminous) objects in the nearby universe. That is the reason that at $z < 0.5$ no SDSS quasars have a black-hole mass above about $10^9$~\Msol{} at $z \approx 0$, while the BQS has a number of objects above this mass. Nonetheless, even for the bright quasar samples of BQS and LBQS, one can see a (rapid) decline in black-hole mass from a redshift of 0.5 to the present. This does not mean, of course, that black holes with mass above 10$^9$~\Msol{} do not exist in the local universe; the giant elliptical galaxy M87 is known to have a black-hole mass of a few billion solar masses. It simply shows that the activity of the most massive black holes rapidly declines from $z = 0.5$ to the present. They simply ``turn off'' and become increasingly quiescent.

% FIGURE 3 (a,b,c)
\begin{figure}
\begin{center}
%\hskip -0.25cm
%\hbox{
\includegraphics[height=6.5cm,width=6.5cm,angle=0]{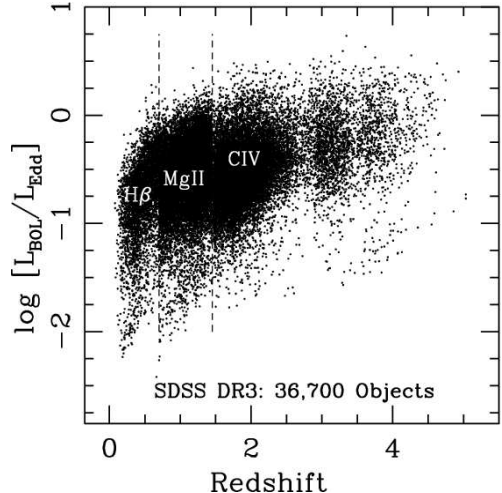}
%\hskip 0.3cm
%\includegraphics[height=5.5cm,width=5.5cm,angle=0]{fig4a.eps}
%}
\end{center}
%\vskip 0.2cm
%\includegraphics[height=6.0cm,width=6.0cm,angle=0]{fig4b.eps}
%\hskip 0.25cm
%\includegraphics[height=6.0cm,width=6.0cm,angle=0]{fig4c.eps}
\caption{
%{\it Upper left:} 
Distribution of Eddington luminosity ratios, $L_{\rm Bol}/L_{\rm Edd}$
for the SDSS DR3 quasar catalog. 
%{\it Upper right:} The imposed luminosity range (between the solid lines) for analysis
%of the potential change in Eddington luminosity ratio with redshift.
%{\it Lower left:} The mean and mode of the $L_{\rm Bol}/L_{\rm Edd}$ distributions 
%in the redshift binned subsets. 
%{\it Lower right:} The mean and mode of the FWHM(\civ) distributions in the redshift
%binned subsets.
%In the lower panels, the horizontal bars indicate the redshift bin width, 
%while the vertical bars show the 10\% and 90\% distribution ranges for each subset in 
%the luminosity range shown in the upper right panel.
%These results are preliminary and need to be confirmed by a more thorough study.
}\label{fig:LoLzevol}
\end{figure}

\subsection{The Eddington luminosity ratio distribution with redshift}

With both the black-hole mass and the bolometric luminosities readily available, 
we can infer the Eddington luminosity ratio, \lol, a crude estimate of the black-hole accretion rate. The distribution of Eddington ratios for the SDSS DR3
quasars is shown in Figure~\ref{fig:LoLzevol}; as seen for the luminosity and mass distributions, the lower flux limits define the lower boundary to the distribution. As expected, most of the quasars appear to radiate between 10\% and 100\% of their Eddington luminosities. Even in the low redshift ($z < 0.5$) universe there are many highly accreting quasars.  Although some objects appear to have super-Eddington accretion rates up to a value of 3 (0.5\,dex), this is not likely to be real. For one, the statistical uncertainties are entirely consistent with no significant population of sources with super-Eddington luminosities. In addition, the \lol\ estimate is based on the assumption of Bondi accretion (\ie spherical accretion), which is not very likely to hold for a black hole fed by an accretion disk.  Moreover, the \lol\ estimates are crude and have uncertainties at least of a factor of about 3.5 due to the uncertainties in the mass estimates. 

The distribution of the \lol\ values with redshift shows that at the highest redshifts the quasars must be highly accreting to be detected. As a result, all
known $z \gsim$4 quasars are accreting at or near their Eddington limits.

\section{Mass functions of actively accreting black holes \label{MF.sec}}

% FIGURE 4
\begin{figure}
\centering
\includegraphics[height=5.5cm,width=5.5cm,angle=0]{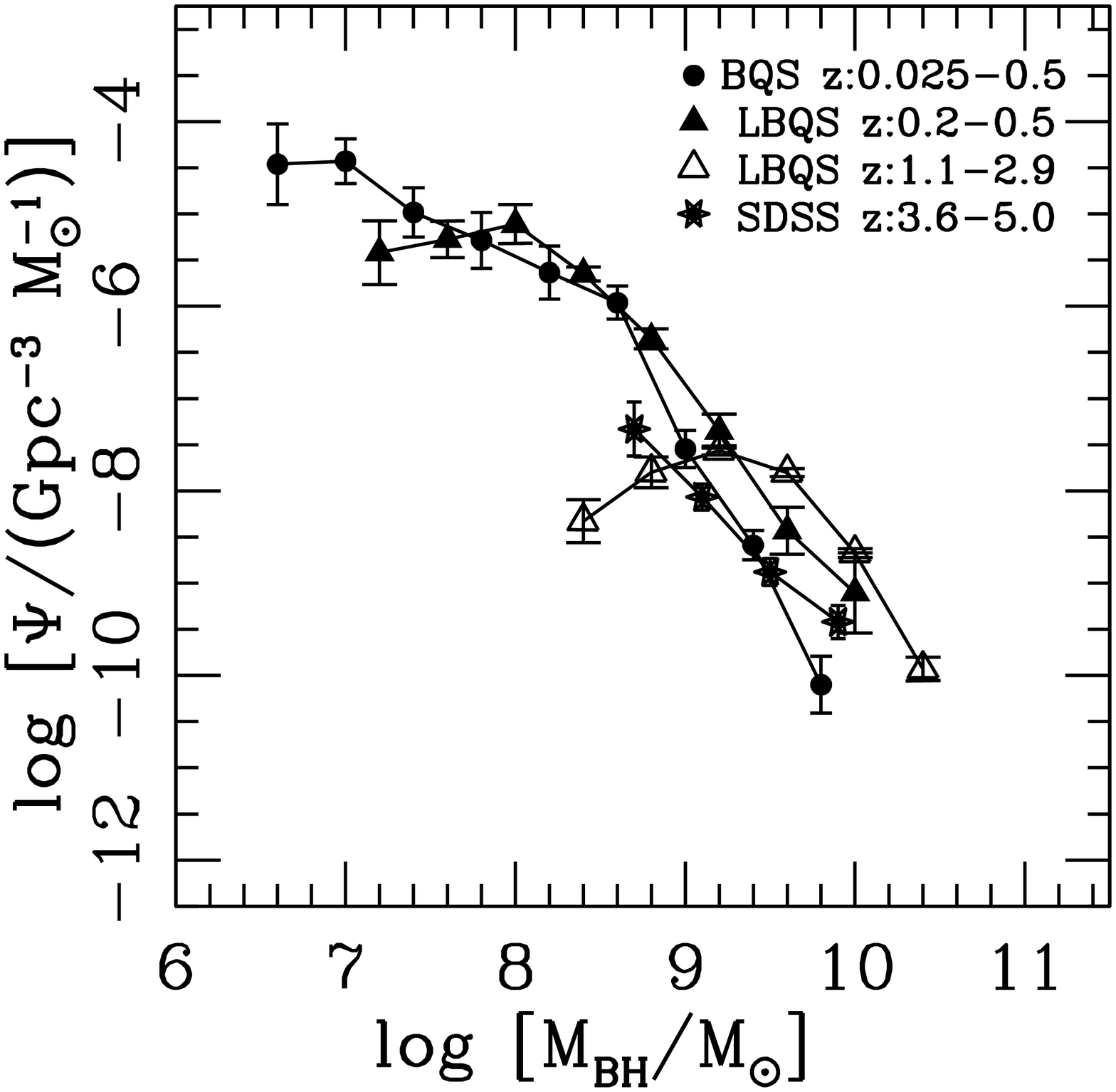}
\hskip 0.4cm
\includegraphics[height=5.5cm,width=5.5cm,angle=0]{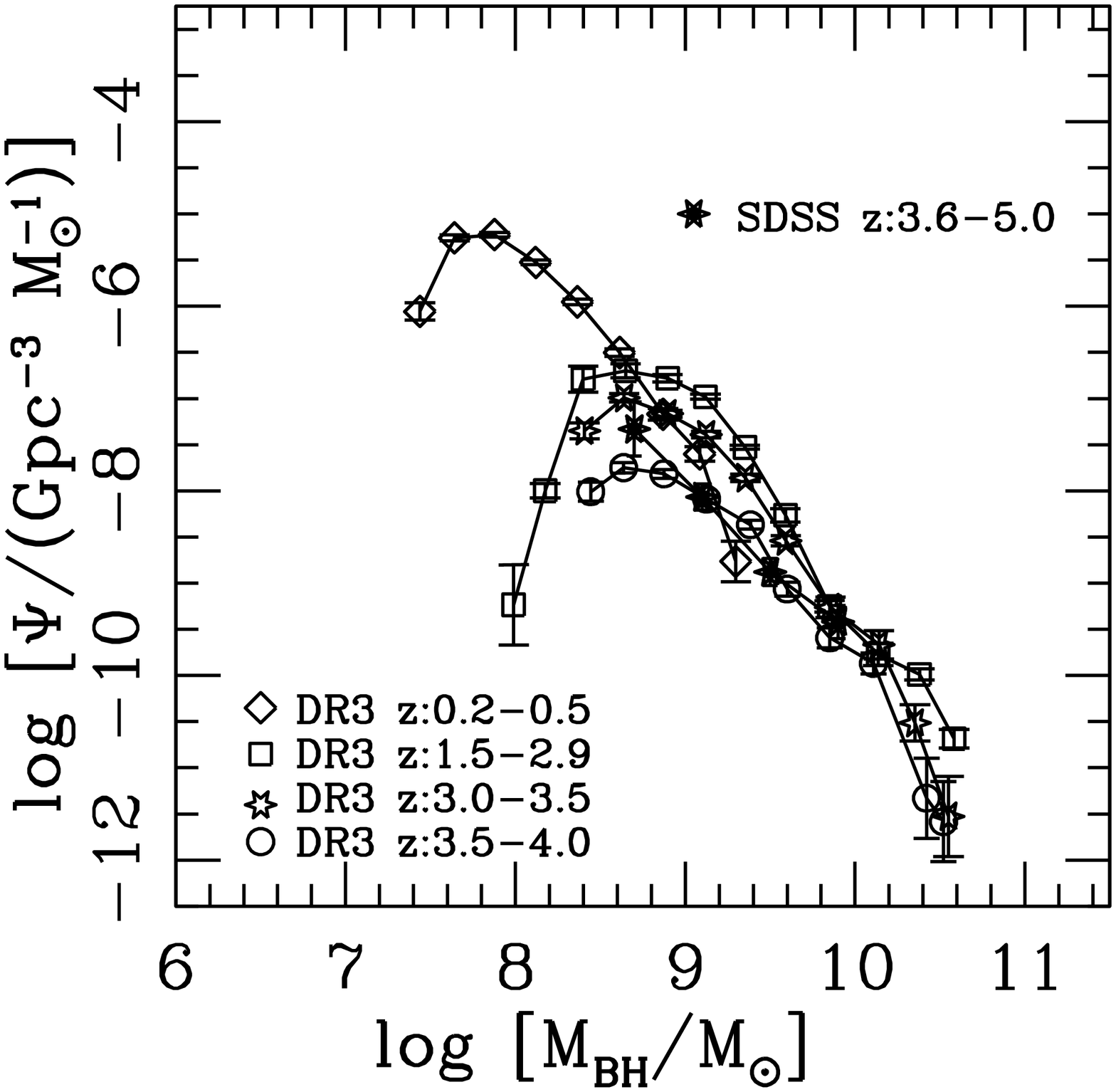}
\vskip 0.2cm
\includegraphics[height=5.5cm,width=5.5cm,angle=0]{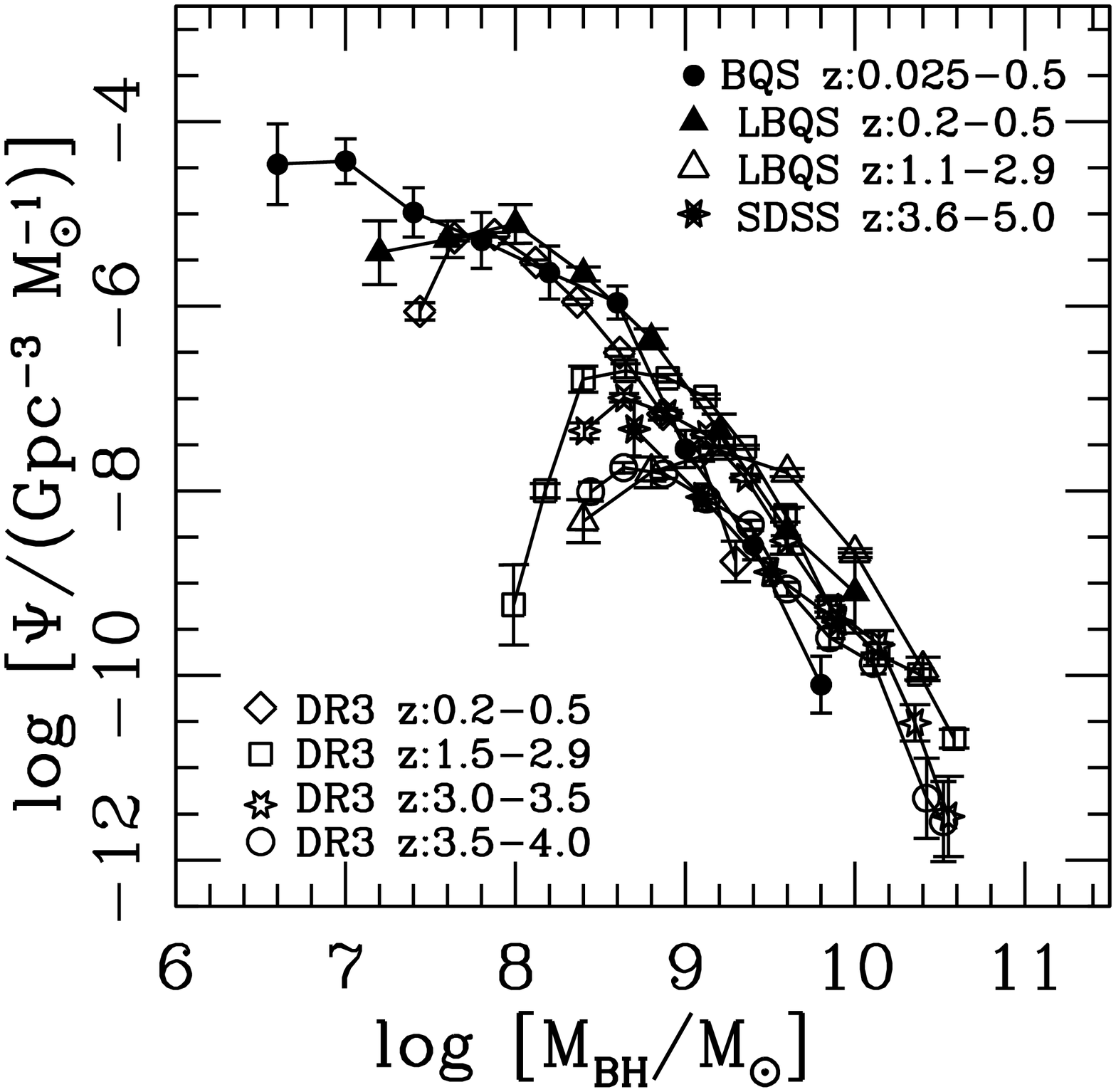}
\caption{Mass functions of actively accreting black holes at different redshifts and
for different quasar samples, as labeled. The ``SDSS'' sample at a redshift of 3.6 to 5.0 is the color-selected sample of Fan \et (2001). The SDSS DR3 mass functions are preliminary. The downturn at the low mass end is not real, but caused by uncorrected incompleteness in these bins. The bottom panel shows how the mass functions shown in the top panels compare to each other.}
\label{fig:MF}
\end{figure}

In the spirit of showing preliminary results from work in progress, I briefly
present and discuss our first cut at the mass functions of actively accreting black holes of quasars at a range of redshifts.  The mass functions for the different quasar samples are shown in Figure~\ref{fig:MF}. Please note that the downturn of the mass functions at the low mass end is an artifact of incompleteness in these bins that is not yet corrected for.

The upper left panel shows the BQS, LBQS, and the color-selected sample of Fan \et (2001). The BQS (filled circles) and low redshift subset of the LBQS (filled triangles) show consistent shape and normalization. The high-$z$ LBQS subset (open triangles) lies slightly higher than its low-$z$ subset (at its high-mass end) and lies even higher than the color-selected sample (filled stars) located at higher redshifts yet.  However, this is entirely consistent with the observed space density of quasars, which peaks at a redshift of about~2 to~3 and declines above and below these redshifts (\eg Peterson 1997). This diagram also seems to
indicate that the comoving volume density of~1 to 10~billion solar mass black 
holes at redshifts of~4 to~5 (the color-selected sample) approximately resembles that of the nearby bright quasars in the BQS (at $z < 0.5$).

The upper right panel shows the mass functions for the SDSS DR3 quasar catalog divided into various redshift bins as labeled. For reference, the color-selected sample from the left panel is also shown. Again, we see a decrease in the normalization of the mass function beyond a redshift of~3. In the bottom panel, the individual mass functions from the top panels are overplotted to show their relative normalizations. In particular, the low-$z$ DR3 subset (open diamonds) is entirely consistent with the other two low-$z$ subsets of the BQS and the LBQS.
A more detailed discussion of the finalized mass functions are presented by Vestergaard \et (2009) and Vestergaard \& Osmer (2009).

\subsection{Constraints on black-hole growth}

\begin{figure}
\centering
\includegraphics[height=9cm,width=12cm,angle=0]{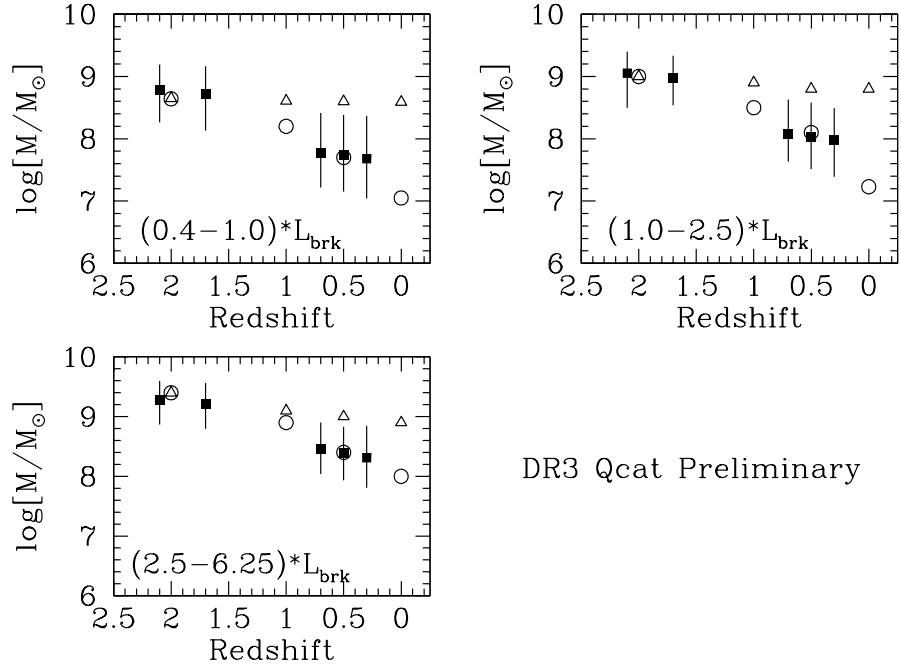}
\caption{Distributions of masses of actively accreting black holes at different
redshifts for the SDSS DR3 quasar catalog. The filled squares in each panel show the observed distribution with redshift for different luminosity ranges, as labeled. The filled squares show the mean value, while the vertical bars show the 10\% to 90\% ranges. The break luminosity, $L_\mathrm{brk}$, is the 
characteristic luminosity at the knee of the quasar luminosity function. Theoretical model predictions are indicated by open symbols: Open triangles: Short accretion phase build-up; Open circles: mass-dependent accretion build-up.}\label{fig:adam}
\end{figure}

The ultimate goal is to use the black-hole mass estimates and the estimates of the mass functions of actively accreting black holes to constrain how black holes grow and evolve over cosmic time. This will be done by comparing these empirical results with theoretical models of how black holes can grow. While a more detailed analysis is yet to be done, it is interesting to do a simple comparison with one such theoretical model. Steed \& Weinberg (2003) modeled how the distribution of black-hole masses change with redshift depending on the dominant growth mechanism. In Figure~\ref{fig:adam} I compare the distribution of quasar black-hole masses in SDSS DR3 with two such growth scenarios from Steed \& Weinberg. The empirical data suggest that the black holes are more likely grow in a fashion in which the accretion rate depends on the black-hole mass rather than one in which the black holes obtain their mass in a short accretion phase, equal to the quasar lifetime. A more detailed discussion will be presented elsewhere.

% FIGURE 5

\subsection{The space density of very massive black holes \label{spacedens.sec}}

In Section~\ref{Mdistrib.sec} and Figure~\ref{fig:Mz} it was shown that the typical black-hole mass at high redshift is a few billion solar masses and extends to 10~billion solar masses. Since the most massive black hole (in the elliptical galaxy M87) measured in the local neighborhood is only about $3\times10^9$~\Msol, and the measurement method does not exclude the determination of more massive black holes, one may question whether the single-epoch mass estimates, discussed here, may indeed be overestimates. I address this issue further in Section~\ref{overestimates.sec} but it is instructive to also look at the high-redshift space density of black holes more massive than M87. Since the mass functions contain information on the comoving space density of the black holes as a function of their mass, I have integrated the current version of the quasar black-hole mass function above a mass of $3\times10^9$~\Msol{} for the subsets with quasars at and above a redshift of 4 (\ie the mass functions shown by the open circles and filled stars in Figure~\ref{fig:MF}). This measurement shows that such massive black holes are so rare that we need to probe the local neighborhood out to a distance of $\sim$290\,Mpc from earth in order to detect just \textit{one} such massive black hole. This is a volume about 25 times larger
than what has been studied at present ($R \approx 100$\,Mpc; Ferrarese \&
Merritt 2000; Tremaine \et 2002). Owing to the uncertainty in the mass estimates, the mass functions may intrinsically be steeper than our estimates in 
Figure~\ref{fig:MF} show. In that case, the space density of the most massive black holes will be even lower, and an even larger local volume is needed. Strictly speaking, based on the uncertainties in the mass estimates, the mass functions could be shallower and thus the frequency of very massive black holes could be higher. We may be able to test this in the future. Figure~1 of Ferrarese (2003) indicates that with a diffraction-limited 8\,m telescope, in principle, it is technically feasible to detect 10$^{10}$~\Msol{} black holes out to distances of about 1500\,Mpc. With a diffraction-limited 30\,m we can possibly probe a volume of several thousand Mpc. 

In conclusion, it seems the explanation for the lack of very massive black holes in the local neighborhood is simply that of their extreme rarity and the relatively small volume that we have probed so far. However, as briefly discussed in Section~\ref{overestimates.sec}, there are now suggestions that the brightest cluster galaxies may have black holes with masses of order $10^{10}$~\Msol. It is, however, still to be clarified whether or not these black holes are the quiescent 
equivalent of the very massive black holes found in distant quasars.

\section{Potential issues with scaling relationships \label{issues.sec}}

There are a few issues that may potentially present problems for the single-epoch mass estimates, including those based on scaling relationships, in the sense that the current estimates may in some sense be skewed or systematically biased. The issues most commonly mentioned in the literature include the effects of the relative inclination of the broad-line region to our line of sight, the potential danger of high-ionization outflows in quasars which may affect the \civ\ line emission, the unexpected narrow \civ\ profile for some active nuclei, and the question whether the single-epoch mass methods may overestimate the actual black-hole mass. 

\subsection{Source inclination issues \label{inclination.sec}}

There are clear indications that the broad emission-line region may have the plane-like geometry because the measured gas velocities are larger for more inclined sources. This has been established for the \hb\ line by Wills \& Browne (1986) and for the base of the \civ\ line by Vestergaard, Wilkes, \& Barthel 
(2000) for radio-loud sources where the ratio of the radio-core flux to the radio-lobe flux yields a crude inclination measure: as the relative inclination of the radio jet increases, the core flux becomes less dominant relative to the radio lobe flux; notably, the direction of the jet is expected to be normal to the central engine's accretion disk. In both studies, the authors saw a lack of objects with 
small inclinations (\ie near face-on) and large line widths. The largest line widths are only seen in the most inclined sources. This ``Zone of Avoidance'' distribution indicates that the gas velocities in a plane parallel or nearly parallel to that of 
the accretion disk are larger than perpendicular to this plane.

Since active nuclei are expected to be randomly oriented (for Type~1 sources: within the inclination ranges that defines this subset: $\lsim\!\!\!45^{\rm o}$; Barthel 1989), and the \mbox{current mass} determinations of active nuclei (both the reverberation-mapping method and mass-scaling relations) do not specifically account for source inclination, inclination effects can contribute to the uncertainties in the black-hole mass of individual active nuclei. For ensemble determinations, this effect is expected to average out owing to the way the mass zero-point was obtained (see Section~\ref{f.sec}).

Source inclination has been suggested to explain the scatter of the reverberation masses around the $M_{\rm BH}$--$\sigma_{\ast}$ relationship (\eg Wu \& Han 2001; Zhang \& Wu 2002). Collin \et (2006) investigated this issue and found that this only appears to work in a statistical sense and is not 
supported by individual sources for which the source inclination is known or well constrained (see Peterson, this volume). However, Collin \et did establish that the FWHM measurement tends to be more sensitive to source inclination than the line dispersion. This is consistent with the finding of Peterson \et (2004) that the line dispersion is the most robust line-width measure for mass determinations. Based on the reverberation-mapping data base, Collin \et provide $f$-factor corrections to mass estimates based on the FWHM(\hb) values if the line-dispersion measurement cannot be obtained. This allows mass estimates for 
which the effects of the relative source inclination is decreased.

However, this correction is currently only available for the \hb\ emission line. Corrections applicable to the \civ\ and \mgii\ emission lines will be addressed in a
future publication.

\subsection{High-ionization outflows \label{outflows.sec}}

There are general concerns that active nuclei typically contain high-ionization outflows that potentially affect the \civ\ line emission, and therefore the mass estimates based thereon. These concerns are based on several observations to suggest this. Leighly (2000) finds for the subset classified as narrow-line Seyfert~1 galaxies (NLS1s), that for increasing nuclear luminosity the \civ\ emission line displays an increasing blue asymmetry. Observations of NLS1s in the far-UV (Yuan \et 2004) support the interpretation that these sources have high-ionization outflows. I\,Zw\,1 is a typical example of a NLS1 with a broad blue asymmetric \civ\ profile. It has a FWHM(\hb) $\approx$~1200~\kms, while its \civ\ line is measured to have FWHM of $\sim$4400~\kms\ (Baskin \& Laor 2005). Since NLS1s typically have \hb\ line widths below 2000~\kms, \civ\ line widths several times that of the \hb\ line are not very representative of the velocities of broad line gas dominated by black-hole gravity. This is the background for the discouragement of using \civ\ for mass estimates for NLS1s, and in general if the \civ\ profile has a triangular and blue asymmetric shape similar to that of I\,Zw\,1 (Vestergaard 2004b).

Because NLS1s are luminous sources with highly accreting (albeit small) black holes, there is a concern that quasars that are also luminous and highly accreting may also display such high-ionization outflows that would make mass estimates based on \civ\ biased or even invalid (\eg Bachev \et 2004; Shemmer \et
2004; Baskin \& Laor 2005). Based on the $z > 1$ samples shown in the upper left panel of Figure~\ref{fig:Mz} I previously established (Vestergaard 2004b) that none of the high-$z$ quasar spectra, at least in those samples, have \civ\ profiles resembling that of I\,Zw\,1. For the sources with the strongest asymmetries, I compared their luminosities and mass estimates with those of the remaining quasars and found no indication at all that the mass estimates were
affected. The mass differences were at the 0.1\,dex level and some subsets with asymmetries even showed a lower mass. Assuming that the quasar samples of that study are typical of the quasar population at those redshifts, it is fair to conclude that high-$z$ quasars with \civ\ line profiles resembling that of I\,Zw\,1 must be rare.

However, it is still possible that the \civ\ emission line is associated with outflowing gas. It is well known that the \civ\ emission line is blueshifted relative to its restframe wavelength of 1549~\AA\ (\eg Wilkes 1984; Espey \et 1989; Tytler \& Fan 1992). Based on about 3800 SDSS quasars, Richards \et (2002) established that as the blueshift increases (up to a maximum of about 2000~\kms), the line-equivalent width decreases, the profile shape changes, and the continuum luminosity increases. When comparing the line profiles of subsets of quasars binned by \civ\ blueshift, the authors concluded that the observed blueshift is generated by an increasing deficit of the red side of the profile. Their Figure~4 does suggest a significant broadening of the \civ\ profile with increasing blueshift. Nonetheless, to determine the effect on the single-epoch mass estimates, a comparison of the FWHM of the profiles with the most and the least blueshift is needed. Based on the entries in Table~2 of Richards \et one can 
infer that, intriguingly, there is only a 15\% difference in the FWHM measurements, which translates into a mere 30\% effect in the mass estimates. Regardless of the cause of the blueshift, since most quasars display a smaller blueshift than 2000~\kms\ and the current mass estimates based on \civ\ scaling relations have uncertainties of order of a factor of~4, the effects of the \civ\ blueshift is insignificant at the present time. However, as the mass estimates improve sufficiently this effect needs to be accounted for.

\subsection{The unexpected relative line widths of \civ and \hb}

Motivated by the blueshifts and potential blue profile asymmetries of \civ, Baskin \& Laor (2005) investigated the \civ- and \hb-profile differences using non-contemporaneous optical- and UV-literature data of the BQS quasar sample. They find, in particular, that for objects with FWHM(\hb) $>$ 4000~\kms, the \civ\ profile is not broader than \hb.  Existing monitoring data and the fact that 
\civ\ is a high-ionization line indicate that \civ\ is emitted from a region closer to the black hole than \hb. The Baskin \& Laor study shows that FWHM(\civ) is not 
always larger than FWHM(\hb), and in fact the former rarely comply with being a factor $\sqrt{2}$ broader, as expected based on the virial relationship, if \hb\ is emitted from a distance twice that of \civ. As it turns out, this issue is much less severe when problem data such as low quality \textit{IUE} data, strongly absorbed profiles, and NLS1s are eliminated from the database (see Vestergaard 
\& Peterson 2006 for details). However, it does remain that FWHM(\civ) does not increase as fast as FWHM(\hb) and an inverse correlation between FWHM(\civ)/FWHM(\hb) and FWHM(\hb) is seen extending to $\pm$0.2\,dex in the FWHM ratio. Part of this difference can be attributed to measurement uncertainties.

How well established is the factor 2 difference in the \civ\ and \hb\ lags? Of the few sources with reliable lags for both lines, three sources (NGC~3783, NGC~5548, and NGC~7469) display \civ\ lags that are about a factor~2 shorter than those of \hb\ for the same epochs (Peterson \et 2004). For NGC~4151 there are no contemporaneous epochs for the two lines, but the data are also consistent with this result (Bentz \et 2006b; Metzroth \et 2006). But for two sources the measured \civ\ and \hb\ lags argue otherwise. Existing data for Fairall~9 display a \civ\ lag that is 1.5 times larger than that of \hb\ and has large errors (Peterson \et 2004). Photoionization theory expects the \civ\ and \lya\ line gas to be emitted from similar distances, so judging from the \lya\ lag, a more reliable \civ\ lag may be at 2/3 the lag of \hb.  For 3C\,390.3 the \civ\ line is narrower than \hb\ and the \civ\ lag is almost twice that of \hb but within the large errors, 
the two lines could be emitted from similar distances (Peterson \et 2004). So judging by the most recent analysis of the reverberation data base of Peterson \et (2004) there are good reasons to expect \hb\ to be emitted from a distance twice that of \civ. But the data on Fairall~9 and 3C~390.3 suggest that perhaps not all sources have broad-line regions that are as neatly organized.
As argued in Section~\ref{newMrel.sec}, the broad-line region is a dynamic place and a neat onion-skin--like emission-line gas distribution is, in fact, not expected. This is supported by photoionization theory. In their Figure~3 Korista \et (1997) present the strength of different emission lines as a function of gas density and incident ionizing luminosity. A comparison of the diagrams for \civ\ and \hb\ shows a significant overlap of parameter space for which sufficiently strong \hb\ and \civ\ can be emitted. Hence, their line regions are not necessarily mutually exclusive, and they can indeed be located closer than the factor~2 difference in distance. In fact, the non-contemporaneous nature of most UV and optical data being compared and the dynamic nature of the broad-line region, which also likely has a clumpy gas distribution, is likely part 
of the explanation for the \civ- and \hb-line widths not always scaling perfectly with one another.

After all, the most compelling argument in favor of \civ-based mass estimates is that reverberation results show the virial products ($R V^2$) are consistent for \textit{all} the measured emission lines, which includes \civ\ (Peterson \& Wandel 1999, 2000; Onken \& Peterson 2002).

\subsection{Are the masses overestimated? \label{overestimates.sec}}

When the scaling relationships are applied to samples of high-redshift quasars,  the estimated typical black-hole mass is of order a billion solar masses and extends to a few tens of billions of solar masses (Section~\ref{Mdistrib.sec}).
The validity of these large black-hole masses have been questioned (\eg Netzer 2003) for a couple of reasons. Firstly, black-hole masses of 10~billion solar masses or more are not found in the local universe among quiescent black holes  assumed to have been active in the past. The most massive black hole detected in the local universe is that in the giant elliptical galaxy, M87, with a mass of 
$3\times10^9$~\Msol{} (Harms \et 1994; Macchetto \et 1997). Second, from the well-established $M_\mathrm{BH}$--$\sigma_{\ast}$ relationship one can infer that a 10~billion solar mass black hole will reside in a galaxy with a bulge that has stellar velocity dispersions above 400~\kms\ (\eg Tremaine \et 2002). Such massive galaxies are not observed in the local universe. It is therefore fair to question whether the single-epoch mass estimates from the scaling relations systematically over-predicts the mass of the black hole by about an order of magnitude. This would bring the highest quasar black-hole mass estimates in full consistency with the black-hole masses in quiescent galaxies measured so far
in the local universe. The methods to determine their black-hole masses are by many considered more reliable since they rely on stellar velocities that are less prone to non-gravitational forces, such as radiation pressure, that is expected to be present and strong in the presumably violent environment near the accreting 
black hole.

There are different considerations supporting that mass-scaling relationships do not systematically overestimate the black-hole mass. A full discussion can be found elsewhere (Vestergaard 2004b). Here I will summarize the main points. First, it is argued in Section~\ref{rlhiz.sec} that the radius--luminosity relationship is fully applicable to high-$z$ quasars. Hence, there is no indication that the use of this relationship should bias us in any way. Second, there is no indication that the kinematics of the broad-line region of high-redshift quasars
are dominated by forces other than black hole gravity. The argument in favor is a combination of the fact that this is certainly found to hold for nearby active nuclei and quasars (Peterson \& Wandel 1999, 2000; Onken \& Peterson 2002; 
Kollatschny 2003) and quasar spectra look very similar at all redshifts (\eg Dietrich
\et 2002). In addition, as noted in Section~\ref{rlhiz.sec}, the monitoring data of the $z \approx 3$ quasar for which Kaspi \et (2007) was able to measure a \civ\ emission-line lag also support the notion that high-$z$ quasar broad-line regions
are very similar to those at lower redshift. Third, it is argued in Section~\ref{outflows.sec} that high-ionization outflows are not important for a very large fraction of quasars. Fourth, one can demonstrate that it is unlikely that the mass
estimates are systematically too large using results from photoionization modeling (\eg Korista \et 1997).  For the sake of this argument, assume an estimated black-hole mass of $10^9$\,\Msol\ that is a factor~10 overestimated. The virial theorem shows that a typical quasar with FWHM(\civ) of 4500~\kms\ will emit the \lya- and \civ-lines from a distance of only $\sim$33 light-days from the central source. At this location for a typical quasar luminosity of \lbol\ of order $10^{47}$~erg~s$^{-1}$, the ratio of photons-to-gas particles is so high that not only are these lines emitted inefficiently, but especially the \ciii-line emission cannot be generated.  The prominence of the \ciii\ in quasar spectra argue for the presence of low-density gas subjected to lower ionizing flux.  The key is that both the \lya\ and the \civ\ lines are much more efficiently emitted from these regions, that necessarily are located much further from the central source in accordance with the predicted distance from the $R$--$L$ relationship.

In conclusion, there are no obvious indications based on our existing knowledge of the broad-line region that the mass estimates based on scaling relations are systematically overestimated. Even the crudeness of using the FWHM of a single-epoch emission line that contains a contribution from non-varying emission line gas does not significantly or systematically affect the mass estimates
(Vestergaard 2004b).

In closing, it is interesting to note that recent findings show that a natural ``saturation'' of the bulge stellar-velocity dispersion occurs above $\sigma_{\ast} \approx 400$~\kms\ (Lauer \et 2007).  For the brightest cluster galaxies and other high $\sigma_{\ast}$ galaxies, $\sigma_{\ast}$ is thus not an accurate indicator of the central black-hole mass. On the other hand, the galaxy luminosity indicates black-hole masses of order 10$^{10}$~\Msol. The apparent contradiction of the large quasar black-hole masses is therefore no longer that evident. Moreover, as I showed in Section~\ref{spacedens.sec}, 10$^{10}$~\Msol{} active black holes are so rare that we should not expect to see any in the local volume that has been probed so far.

\section{Summary and conclusions \label{summary.sec}}

It is argued that mass-scaling relations are the preferred method to estimate black hole masses of distant active galaxies and quasars owing to the ease with
which the method can be applied to large samples, its higher accuracy relative to alternative comparable methods, and the reliability of the mass estimates. Furthermore, the accuracy can potentially be improved. This is important because we seek to use these mass estimates to understand how black holes grow and affect galaxies and their evolution---those that host the active black holes and those in their neighborhood.
At present, the statistical 1$\sigma$ uncertainty in the absolute mass values amounts to a factor of 3.5 to 5, depending on the emission line and line-width measure applied; the \civ\ relations have lower uncertainties.

Distributions of black~hole masses of distant quasars at a range of epochs is presented.  The estimated masses are very large, a billion \Msol{} or more, even at redshifts of 4 to 6; the black holes can clearly grow and mature very quickly in the early universe. However, a maximum mass of $10^{10}$~\Msol{} is also observed, which shows that black holes reach their ultimate growth limit then and thus must shut off their activity. 
Preliminary black-hole mass functions of various large quasar samples is also presented and discussed; a full analysis will be presented elsewhere (Vestergaard \et 2008; Vestergaard \& Osmer 2009; Vestergaard \et 2009, in preparation). Furthermore, I outlined and briefly discussed some potential issues associated with applying mass-scaling relations to luminous high-redshift quasars and concluded that none of these jeopardize this
method or significantly affects the mass estimates. Also, there are no indications that the mass estimates are systematically too high.

\begin{acknowledgments}
I am grateful for financial support of this work by NASA through grants HST-AR-10691 and HST-GO-10417 from the Space Telescope Science Institute, and by NSF grant AST-0307384 to the University of Arizona. I thank Xiaohui Fan and Bradley Peterson for discussions related to this work.
\end{acknowledgments}


\begin{thebibliography}{}

\bibitem[Bachev et al. 2004]{}\textsc{Bachev, R., et~al.} 2004 \textit{ApJ} \textbf{617}, 171.
\bibitem[Baldwin et al. 1995]{1995ApJ...455L.119B}\textsc{Baldwin, J., et al.} 1995 \textit{ApJ} \textbf{455}, L119.
\bibitem[Barth et al. 2003]{}\textsc{Barth, A., et al.} 2003 \textit{ApJ} \textbf{594}, L95.
\bibitem[Barth et al. 2005]{2005ApJ...619L.151B}\textsc{Barth, A., et al.} 2005 \textit{ApJ} \textbf{619}, L151.
\bibitem[Barthel 1989]{1989ApJ...336..606B}\textsc{Barthel, P.~D.} 1989 \textit{ApJ} \textbf{336}, 606.
\bibitem[Baskin and Laor 2005]{}\textsc{Baskin, A.\ \& Laor, A.} 2005 \textit{MNRAS} \textbf{356}, 1029.
\bibitem[Bentz et al. 2006a]{2006ApJ...644..133B}\textsc{Bentz, M., et al.} 2006a \textit{ApJ} \textbf{644}, 133.
\bibitem[Bentz et al. 2006b]{2006ApJ...651..775B}\textsc{Bentz, M., et al.} 2006b \textit{ApJ} \textbf{651}, 775.
\bibitem[Boroson 2003]{}\textsc{Boroson, T.~A.} 2003 \textit{ApJ} \textbf{585}, 647.
\bibitem[Collin et al. 2006]{}\textsc{Collin, S., et~al.} 2006 \textit{A\&A} \textbf{456}, 75.
\bibitem[Dietrich et al. 2002]{2002ApJ...581..912D}\textsc{Dietrich, M., et al.} 2002 \textit{ApJ} \textbf{581}, 912.
\bibitem[Dietrich et al. 2003a]{}\textsc{Dietrich, M., et al.} 2003a \textit{ApJ} \textbf{589}, 722.
\bibitem[Dietrich et al. 2003b]{}\textsc{Dietrich, M., et al.} 2003b \textit{ApJ} \textbf{596}, 817.
\bibitem[Dietrich and Hamann 2004]{}\textsc{Dietrich, M.\ \& Hamann, F.} 2004 \textit{ApJ} \textbf{611}, 761.
\bibitem[Djorgovski and Davis 1987]{1987ApJ...313...59D}\textsc{Djorgovski, S.\ \& Davis, M.} 1987 \textit{ApJ} \textbf{313}, 59.
\bibitem[Dressler et al. 1987]{1987ApJ...313...42D}\textsc{Dressler, A., et al.} 1987 \textit{ApJ} \textbf{313}, 42.
\bibitem[Dwek et al. 2007]{2007ApJ...662..927D}\textsc{Dwek, E., et~al.} 2007 \textit{ApJ} \textbf{662}, 927.
\bibitem[Elvis et al. 1994]{1994ApJS...95....1E}\textsc{Elvis, M., et al.} 1994 \textit{ApJS} \textbf{95}, 1.
\bibitem[Espey et al. 1989]{}\textsc{Espey, B., et al.} 1989 \textit{ApJ} \textbf{342}, 666.
\bibitem[Falomo et al. 2004]{2004cbhg.sympE..14F}\textsc{Falomo, R., et al.} 2004. In \textit{Carnegie Observatories Astrophysics Series, Vol.~1: Coevolution of Black Holes and Galaxies} (ed.\ L.~C.\ Ho). Carnegie Observatories,\hfil\break http://www.ociw.edu/ociw/symposia/series/symposium1/proceedings.html/.
\bibitem[Fan et al. 2001]{2001AJ....121...31F}\textsc{Fan, X., et al.} 2001 \textit{AJ} \textbf{121}, 31.
\bibitem[Ferrarese 2003]{}\textsc{Ferrarese, L.} 2003. In \textit{Hubble's Science Legacy: Future Optical-Ultraviolet Astronomy from Space} (eds.\ K.~R.\  Sembach, J.~C.\ Blades, G.~D.\ Illingworth, \&R.~C.\ Kennicutt, Jr.). ASP Conference Series, Vol.~291, p.~196. ASP.
\bibitem[Ferrarese and Merritt 2000]{}\textsc{Ferrarese, L.\ \& Merritt, D.} 2000 \textit{ApJ} {\bf 539}, L9.
\bibitem[Ferrarese et al. 2001]{2001ApJ...555L..79F}\textsc{Ferrarese, L., et~al. } 2001 \textit{ApJ} {\bf 555}, L79.
\bibitem[Fria\c{c}a and Terlevich 1998]{}\textsc{Fria\c{c}a, A.\ \& Terlevich, R.} 1998 \textit{MNRAS} \textbf{298}, 399.
\bibitem[Gebhardt et al. 2000a]{2000ApJ...539L..13G}\textsc{Gebhardt, K., et al.} 2000a \textit{ApJ} \textbf{539}, L13.
\bibitem[Gebhardt et al. 2000b]{2000ApJ...543L...5G}\textsc{Gebhardt, K., et al.} 2000b \textit{ApJ} \textbf{543}, L5.
\bibitem[Harms et al. 1994]{}\textsc{Harms, R.~J., et al.} 1994 {\it ApJ} {\bf 435}, L35.
\bibitem[Hewett et al. 1995]{}\textsc{Hewett, P., et al.} 1995 \textit{AJ} \textbf{109}, 1498.
\bibitem[Jiang et al. 2007]{2007arXiv0707.1663J}\textsc{Jiang, L., et~al.} 2007 \textit{ApJ} \textbf{134}, 1150.% in press (astro-ph/0707.1663)
\bibitem[Kaspi 2001]{2001ASPC..224..347K}\textsc{Kaspi, S.} 2001. In \textit{Probing the Physics of Active Galactic Nuclei} (eds.\ B.~M.\ Peterson, R.~W.\  Pogge, \& R.~S.\ Polidan). ASP Conference Proceedings, Vol.~224, p.~347. ASP.
\bibitem[Kaspi et al. 2000]{}\textsc{Kaspi, S., et al.} 2000 \textit{ApJ} \textbf{533}, 631.
\bibitem[Kaspi et al. 2005]{2005ApJ...629...61K}\textsc{Kaspi, S., et al.} 2005 \textit{ApJ} \textbf{629}, 61.
\bibitem[Kaspi et al. 2007]{2007ApJ...659...997K}\textsc{Kaspi, S., et al.} 2007 \textit{ApJ} \textbf{659}, 997.
\bibitem[Kollatschny 2003]{}\textsc{Kollatschny, W. } 2003 \textit{A\&A} \textbf{407}, 461.
\bibitem[Korista et al. 1997]{}\textsc{Korista, K., et al.} 1997 \textit{ApJS} \textbf{108}, 401.
\bibitem[Kukula et al. 2001]{2001MNRAS.326.1533K}\textsc{Kukula, M., et al.} 2001 \textit{MNRAS} \textbf{326}, 1533.
\bibitem[Laor 2000]{}\textsc{Laor, A.} 2000 \textit{ApJ} \textbf{543 }, L111.
\bibitem[Lauer et al. 2007]{}\textsc{Lauer, T., et al.} 2007 \textit{ApJ} \textbf{662}, 808.
\bibitem[Leighly 2000]{}\textsc{Leighly, K.} 2000 \textit{NewAR} \textbf{44}, 395.
\bibitem[Macchetto et al. 1997]{}\textsc{Macchetto, D., et al.} 1997 {\it ApJ} \textbf{489}, 579.
\bibitem[Magorrian et al. 1998]{1998AJ....115.2285M}\textsc{Magorrian, J., et al.} 1998 \textit{AJ} \textbf{115}, 2285.
\bibitem[Maiolino et al.\ 2005]{2005A&A...440L..51M}\textsc{Maiolino, R., et~al.} 2005 \textit{A\&A} \textbf{440}, 51.
\bibitem[Matteucci and Recchi 2001]{}\textsc{Matteucci, F.\ \& Recchi, S.}  2001 \textit{ApJ} \textbf{558}, 351.
\bibitem[Metzroth et al. 2006]{}\textsc{Metzroth, K.~G., Onken, C.~A., \& Peterson, B.~M.} 2006 \textit{ApJ} \textbf{647}, 901.
\bibitem[McLure and Jarvis 2002]{2002MNRAS.337..109M}\textsc{McLure, R.\ \& Jarvis, M.} 2002 \textit{MNRAS} \textbf{337}, 109.
\bibitem[McLure and Dunlop 2002]{2002MNRAS.331..795M}\textsc{McLure, R.\ \& Dunlop, J.} 2002 \textit{MNRAS} \textbf{331}, 795.
\bibitem[Nelson 2000]{}\textsc{Nelson C.} 2000 \textit{ApJ} \textbf{544}, L91.
\bibitem[Nelson and Whittle 1996]{}\textsc{Nelson, C.\ \& Whittle, M.} 1996 \textit{ApJ} \textbf{465}, 96.
\bibitem[Netzer 2003]{}\textsc{Netzer, H.} 2003 \textit{ApJ} \textbf{583}, L5.
\bibitem[Onken and Peterson 2002]{2002ApJ...572..746O}\textsc{Onken, C.~A.\ \& Peterson, B.~M.} 2002 \textit{ApJ} \textbf{572}, 746.
\bibitem[Onken et al. 2004]{2004ApJ...615..645O}\textsc{Onken, C.~A., et al.} 2004 \textit{ApJ} \textbf{615}, 645.
\bibitem[Peterson 1993]{1993PASP..105..247P}\textsc{Peterson, B.~M.} 1993 \textit{PASP} \textbf{105}, 247.
\bibitem[Peterson 1997]{}\textsc{Peterson, B.~M.} 1997 \textit{An Introduction to Active Galactic Nuclei}. Cambridge University Press.
\bibitem[Peterson and Wandel 1999]{}\textsc{Peterson, B.~M.\ \& Wandel, A.} 1999 \textit{ApJ} \textbf{521}, L95.
\bibitem[Peterson and Wandel 2000]{}\textsc{Peterson, B.~M.\ \& Wandel, A.} 2000 \textit{ApJ} \textbf{540}, L13.
\bibitem[Peterson et al. 2004]{2004ApJ...613..682P}\textsc{Peterson, B.~M., et al.} 2004 \textit{ApJ} \textbf{613}, 682.
\bibitem[Peterson et al. 2005]{2005ApJ...633..799P}\textsc{Peterson, B.~M., et al.} 2005 \textit{ApJ} \textbf{633}, 799.
\bibitem[Richards et al. 2002]{2002AJ....124....1R}\textsc{Richards, G.~T., et al.} 2002 \textit{AJ} \textbf{124}, 1.
\bibitem[Richards et al. 2006]{2006APJS..166..470R}\textsc{Richards, G.~T., et al.} 2006 \textit{ApJS} \textbf{166}, 470.
\bibitem[Shemmer et al. 2004]{}\textsc{Shemmer, O., et~al.} 2004 \textit{ApJ} \textbf{614}, 547.
\bibitem[Schmidt and Green 1983]{}\textsc{Schmidt, M.\ \& Green, R.~F.} 1983 \textit{ApJ} \textbf{269}, 352.
\bibitem[Schneider et al. 2005]{}\textsc{Schneider, D., et al.} 2005 \textit{AJ} \textbf{130}, 367.
\bibitem[Steed and Weinberg 2003]{2003astro.ph.11312S}\textsc{Steed, A.\ \& Weinberg, D.~H.} 2003; astro-ph/0311312.
\bibitem[Tremaine et al. 2002]{2002ApJ...574..740T}\textsc{Tremaine, S., et~al.} 2002 \textit{ApJ} \textbf{574}, 740.
\bibitem[Tytler and Fan 1992]{}\textsc{Tytler, D.\ \& Fan, X.} 1992 \textit{ApJS} \textbf{79}, 1
\bibitem[Veron et al. 2004]{2004A&A...417..515V}\textsc{Veron, M.-P., et al.} 2004 \textit{A\&A} \textbf{417}, 515.
\bibitem[Vestergaard 2000]{2000PASP..112.1504V}\textsc{Vestergaard, M.} 2000, \textit{PASP} \textbf{112}, 1504.
\bibitem[Vestergaard 2002]{}\textsc{Vestergaard, M.} 2002 \textit{ApJ} \textbf{571}, 733.
\bibitem[Vestergaard 2003]{2003ApJ...599..116V}\textsc{Vestergaard, M.} 2003 \textit{ApJ} \textbf{599}, 116.
\bibitem[Vestergaard 2004a]{}\textsc{Vestergaard, M.} 2004a. In \textit{AGN Physics with the Sloan Digital Sky Survey} (eds.\ G.~T.\ Richards \& P.~B.\ Hall). ASP Conference Series, Vol.~311, p.~69. ASP.
\bibitem[Vestergaard 2004b]{}\textsc{Vestergaard, M.} 2004b \textit{ApJ} \textbf{601}, 676.
\bibitem[Vestergaard et al. 2008]{}\textsc{Vestergaard, M., Fan, X., Tremonti, C.A., Osmer, P.S., Richards, G.T.} 2008 \textit{ApJ} \textbf{674}, L1.
\bibitem[Vestergaard and Osmer 2009]{}\textsc{Vestergaard, M.\ \& Osmer, P.S.} 2009 \textit{submitted to ApJ}
\bibitem[Vestergaard and Peterson 2006]{}\textsc{Vestergaard, M.\ \& Peterson, B.M.} 2006 \textit{ApJ} \textbf{641}, 689.
\bibitem[Vestergaard and Wilkes 2001]{}\textsc{Vestergaard, M.\ \& Wilkes, B.~J.} 2001 \textit{ApJS} \textbf{34}, 1.
\bibitem[Vestergaard et al. 2000]{}\textsc{Vestergaard, M., Wilkes, B.~J., \&  Barthel, P.~D.} 2000 \textit{ApJ} \textbf{538}, L103.
\bibitem[Wandel 2002]{}\textsc{Wandel, A.} 2002 \textit{ApJ} \textbf{565}, 762.
\bibitem[Wandel, Peterson, Malkan 1999]{}\textsc{Wandel, A., Peterson, B.~M., Malkan, M.} 1999 \textit{ApJ} \textbf{526}, 579.
\bibitem[Warner et al. 2003]{2003ApJ...596...72W}\textsc{Warner, C., et~al.} 2003 \textit{ApJ} \textbf{596}, 72.
\bibitem[Wilkes 1984]{}\textsc{Wilkes, B.~J.} 1984 \textit{MNRAS} \textbf{207}, 73.
\bibitem[Wills and Browne 1986]{}\textsc{Wills, B.\ \& Browne, I.} 1986 \textit{ApJ} \textbf{302}, 56.
\bibitem[Woo and Urry 2002]{}\textsc{Woo, J.-H.\ \& Urry, C.~M.} 2002 \textit{ApJ} \textbf{579}, 530.
\bibitem[Wu and Han 2001]{}\textsc{Wu, X.\ \& Han, J.~L.} 2001 \textit{ApJ} \textbf{561}, L59.
\bibitem[Wu et al. 2004]{}\textsc{Wu, X.-B., et~al.} 2004 \textit{A\&A} \textbf{424}, 793.
\bibitem[Yuan et al.(2004)]{2004IAUS..217..364Y} \textsc{Yuan, Q., Brotherton, M., Green, R.~F., \& Kriss, G.~A.} 2004. In \textit{Recycling Intergalactic and Interstellar Matter} (eds.\ P.-A.\ Duc, J.~Braine, \& E.~Brinks). IAU Symposium Series,  Vol.~217, p.~364. ASP.
\bibitem[Zhang and Wu 2002]{}\textsc{Zhang, T.-Z.\ \& Wu, X.-B.} 2002 \textit{Chin.~J.\ Astron.\ Astrophys.} \textbf{2}, 487.

\end{thebibliography}
\end{document}